\documentclass[12pt]{article}

\usepackage[a4paper, total={6in, 8in}]{geometry}
\usepackage[utf8]{inputenc}

\usepackage{longtable}

\usepackage{lineno}

\usepackage{authblk}
\usepackage{setspace}
\usepackage{array}
\usepackage[backend=biber,style=nature]{biblatex}
\usepackage[colorlinks=true]{hyperref}
\usepackage{graphicx}
\usepackage{amssymb}
\usepackage{amsmath}
\usepackage{setspace}
\usepackage{color}
\usepackage{url}
\usepackage{totcount}
\usepackage{lscape}
\usepackage{booktabs}
\usepackage[scaled=.90]{helvet}
\usepackage{caption}
\usepackage{eqnarray}

\DeclareCaptionFont{xipt}{\fontsize{9}{11}\sf\selectfont}
\newcommand{\filename}{main}
\newcommand{\wordcount}[1]{
    \immediate\write18{texcount -sub=none -sum=1,1,1,1,1,1 -merge \filename.tex  | grep "#1" | sed -e 's/+.*//' > \jobname.wcdetail}
\input{\jobname.wcdetail}}

\graphicspath{{new_figs/,figs,SI_figs/}}

\title{Quantifying the Impact of Biobanks and Cohort Studies}

\author[1,2]{Rodrigo Dorantes-Gilardi}
\author[1,5,6]{Kerry Ivey}
\author[1]{Lauren Costa}
\author[1]{Rachael Matty}
\author[1,5,6]{Kelly Cho}
\author[1,5,6,7]{John Michael Gaziano}
\author[1,2,3,4*]{Albert-L\'aszl\'o Barab\'asi}

\affil[1]{Million Veteran Program (MVP) Coordinating Center, VA Boston Healthcare System, Boston, MA, USA}
\affil[2]{Network Science Institute, Northeastern University, Boston, USA}
\affil[3]{Department of Medicine, Brigham and Women's Hospital, Harvard Medical School, Boston, USA}
\affil[4]{Department of Data and Network Science, Central Eastern University, Hungary}
\affil[5]{Division of Aging, Department of Medicine, Brigham and Women’s Hospital, Boston, MA}
\affil[6]{Department of Medicine, Harvard Medical School, Boston, MA}
\affil[7]{Division of General Internal Medicine, Department of Medicine, Brigham and Women’s Hospital, Boston, MA}
\affil[*]{barabasi@gmail.com}

\date{}

\begin{document}

\newtotcounter{citenum} 
\def\oldbibitem{} \let\oldbibitem=\bibitem
\def\bibitem{\stepcounter{citenum}\oldbibitem}

\maketitle



\pagebreak

\begin{abstract}

Biobanks advance biomedical and clinical research by collecting and offering data and biological samples for numerous studies. However, the impact of these repositories varies greatly due to differences in their purpose, scope, governance, and data collected. Here, we computationally identified 2,663 biobanks and their textual mentions in 228,761 scientific articles, 16,210 grants, 15,469 patents, 1,769 clinical trials, and 9,468 public policy documents, helping characterize the academic communities that utilize and support them. We found a strong concentration of biobank-related research on a few diseases, where 20\% of publications focus on obesity, Alzheimer's disease, breast cancer, and diabetes. Moreover, collaboration, rather than citation count, shapes the community's recognition of a biobank. We show that, on average, 41.1\% of articles miss to reference any of the biobank's reference papers and 59.6\% include a biobank member as a co-author. Using a generalized linear model, we identified the key factors that contribute to the impact of a biobank, finding that an impactful biobank tends to be more open to external researchers, and that quality data---especially linked medical records---as opposed to large data, correlates with a higher impact in science, innovation, and disease. The collected data and findings are accessible through an open-access web application intended to inform strategies to expand access and maximize the value of these valuable resources.  
\end{abstract}

\begin{refsection}

\section*{Introduction}
In 2009, Time magazine listed biobanks among the ten ideas changing the world~\cite{park_10_2009}. Indeed, biobanks, repositories of biological samples and curated data pertaining to individuals and disease, are fundamental resources for biomedical research~\cite{kinkorova_biobanks_2016, wichmann_comprehensive_2011, shilo_axes_2020}, being supported by both public and private research centers and funding agencies, as well as the research community at large~\cite{caulfield_review_2014,kinkorova_biobanks_2018}.

One of the first biobanks, The Framingham Heart Study (FHS), was established as a cohort study in 1948 to document the health of 5,209 adult residents from Framingham, Massachusetts, helping define the models still used today for cardiovascular and heart disease risk prediction~\cite{kannel_factors_1961,hajar_framingham_2016}. Currently monitoring the third offspring of the original cohort, FHS is one of the most influential resources in modern medicine~\cite{lloyd-jones_framingham_2004, mahmood_framingham_2014}. Equally influential is a relatively new UK Biobank, founded in 2006 to collect the genetic information, lifestyle, diet, and medical records of 500,000 adults from the United Kingdom~\cite{bycroft_uk_2018}. The datasets arising from the UK Biobank are widely used to advance the understanding of the genetic basis of disease, genetic epidemiology, and public health~\cite{turro_whole-genome_2020,warrington_maternal_2019,ruth_using_2020,griffith_collider_2020}. Following the release of these results,  biobanks like the Million Veteran Program in the United States~\cite{gaziano_million_2016}, or the Kadoorie Biobank in China~\cite{chen_china_2011}, and the HUNT biobank in Norway~\cite{krokstad_cohort_2013,brumpton_hunt_2022}, among others, have been established to link participants' genetic information to their electronic health records.

Although much attention has focused on the FHS and the UK Biobank, these highly visible resources coexist with thousands of biobanks worldwide~\cite{odonoghue_how_2021}. Quantifying and understanding the scientific impact of biobanks is a challenging task, given the great diversity in their goals, usage policies, and cohort characteristics. As a result, currently, we lack a summary-level understanding of the breadth and the diversity of biobanks, nor do we understand how biobanks individually and collectively have impacted our understanding of human health and disease. Here, we fill this gap by relying on big data and the tools of Science of Science~\cite{fortunato_science_2018,wang_science_2021,foster_tradition_2015,fortunato_community_2010} to identify, catalog, and analyze the characteristics of 2,663 biobanks, mapping out all the publications, grants, patents, clinical trials, and public policy documents where these resources are textually mentioned. We use this dataset to track the research footprint of each biobank, offering the first quantitative analysis of biobank usage and impact across multiple dimensions, including research, innovation, public policy, and disease.

We show that biobank-related research over-represents certain disease areas like aging and obesity, while under-representing others like immunotherapy or rare diseases. Moreover, we find that the true impact of biobanks is not visible to the traditional measures of scientific impact. Indeed, On average, only 58.8\% of papers mentioning a biobank cite its reference articles, reducing measurable impact. Moreover, we find that multidimensional impact of biobanks is better tracked through sources like patents, clinical trials, and policy documents, which often do not correlate with their publication counts.  
To measure the true impact of a biobank, we introduce the  Biobank Impact Factor (BIF), tracking a biobank's impact across research \& funding, patent applications, clinical trials, and public health initiatives. We then develop a generalized linear model to uncover the features that explain the visibility and impact of a biobank, finding that biobanks with reputable scientific leaders, open data access for external researchers, and diverse genetic data like genome sequencing or gene-environment interactions tend to have a larger impact.

\subsection*{The Dataset}

Biobanks and cohort studies are closely intertwined, often leading to interchangeable use. Typically, a cohort study is initiated along an associated biobank, maintaining the study's original name. Conversely, some biobanks are designed to support multiple cohort studies, showcasing their versatile application in research. Here, we use the term biobank independently from the process and history responsible to its creation. Estimating the total number of biobanks worldwide is challenging due to the lack of a comprehensive, centralized data source. A recent study estimated around 2 biobanks per 1 million people globally, suggesting approximately 660 biobanks in the USA alone~\cite{odonoghue_how_2021}. Other estimates place the global number in the realm of just a few hundred biobanks~\cite{illumina_inc_population_2020}. The existing biobank directories are often limited in scope, covering specific regions or research projects with varying methodologies~\cite{swertz_towards_2022}. Further complicating biobank tracking is the issue of name multiplicity, where typing errors or references to secondary collections lead to inconsistent naming conventions~\cite{mabile_quantifying_2013}. As well as inconsistencies on what constitutes a biobank~\cite{annaratone_basic_2021}. Despite these challenges, attempts to quantify and catalog biobanks remain crucial for assessing their overall impact on research and healthcare.


Based on the definition of a biobank as a ``large collection of human biological material linked to relevant personal and health information''~\cite{otlowski_biobanks_2020,annaratone_basic_2021}, our dataset includes resources that provide physical or digital human biological data associated with lifestyle, demographic, or health information, such as cohort studies, cancer registries, and large surveys with biological data, as well as tissue, blood, and brain banks.

To identify the true corpus of biobanks, we integrated 16 biobank catalogs and expanded this list by systematically scanning the full scientific literature for mentions of human biobanks (SI Section~\ref{SI:biobank_dataset}). We also filtered all documents (academic publications, patents, grants, clinical trials, and policy documents) mentioning a biobank in the Dimensions relational database~\cite{hook_dimensions_2018}. Finally, we employed natural language processing and co-citation similarity techniques to remove duplicate biobank entries (SI Section~\ref{SI:biobank_mentions}).

\begin{figure}[h]
    \centering
    \includegraphics[width=1\linewidth]{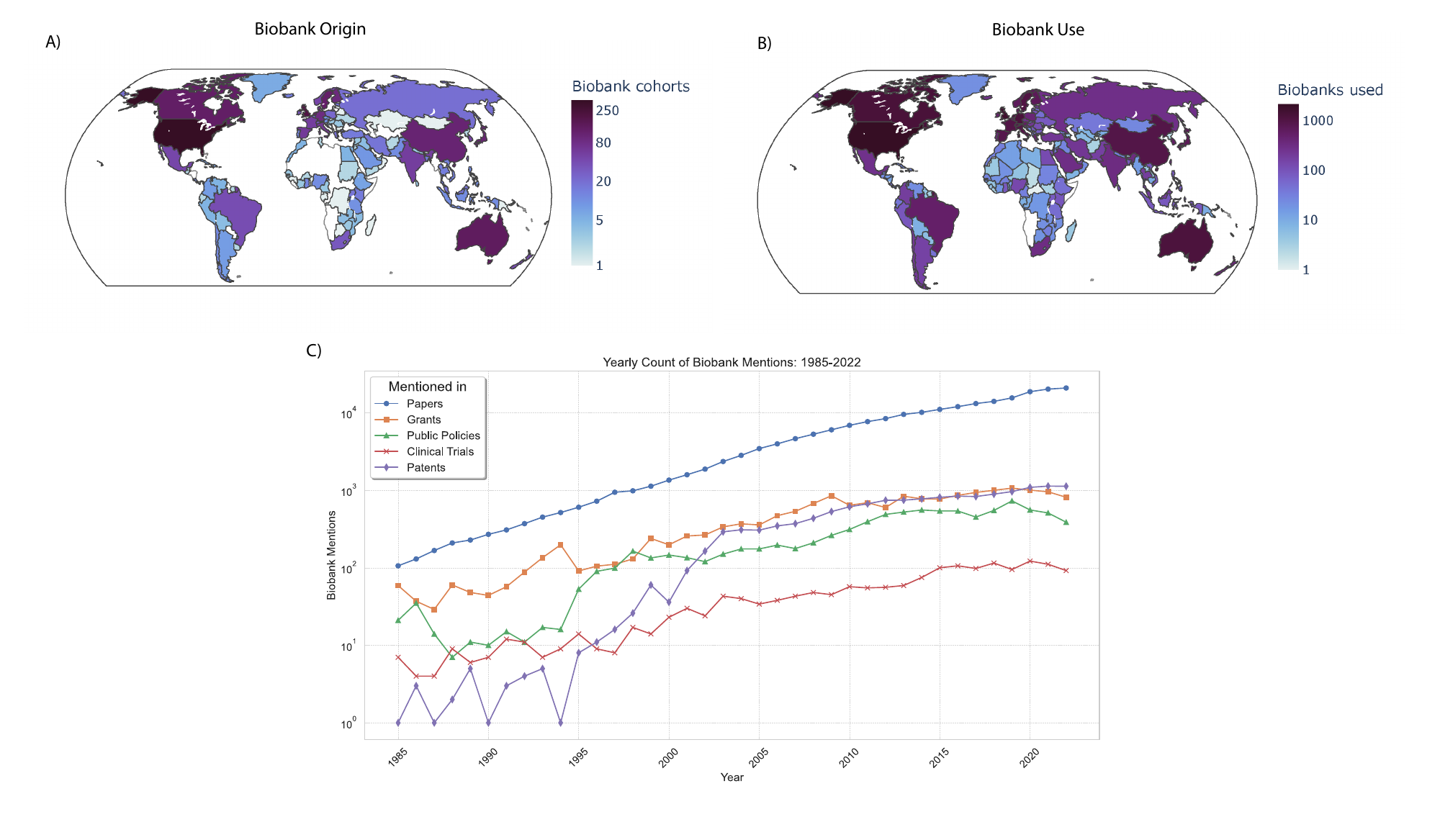}
    \caption{\textbf{Biobank origin, use, and mentions}. A) The origin of a biobank is based on the nationalities of the biobank's cohort sample (SI~\ref{SI:biobank-data}). B) The countries using a biobank are based on the affiliation of authors mentioning a biobank in their publications. D) Year distribution of biobank mentions across papers, grants, patents, clinical trials, and public policy documents between 1985 and 2022.}
    \label{fig:mentions}
\end{figure}

Through this computational approach, we identified 2,663 unique biobanks that originated from and were utilized in 139 countries (Figure~\ref{fig:mentions}A,B). These biobanks were mentioned across 228,761 scientific articles, 16,210 grants, 1,769 clinical trials, 15,469 patents, and 9,468 public policy documents (Figure~\ref{fig:mentions}C). Based on these documents, we extracted multiple features related to the biobank's cohort composition, data offered, and its overall impact (SI~\ref{SI:biobank-data} and Table~\ref{tab:biobank_features}).

The full dataset, as well as the code to produce it, are available at \url{github.com/Barabasi-Lab/quantifying_biobanks} and \url{10.5281/zenodo.11671294}.


\section*{Results}

\subsection*{The Disease Focus of Biobanks}

\begin{figure}[h]
    \centering
    \includegraphics[width=1\linewidth]{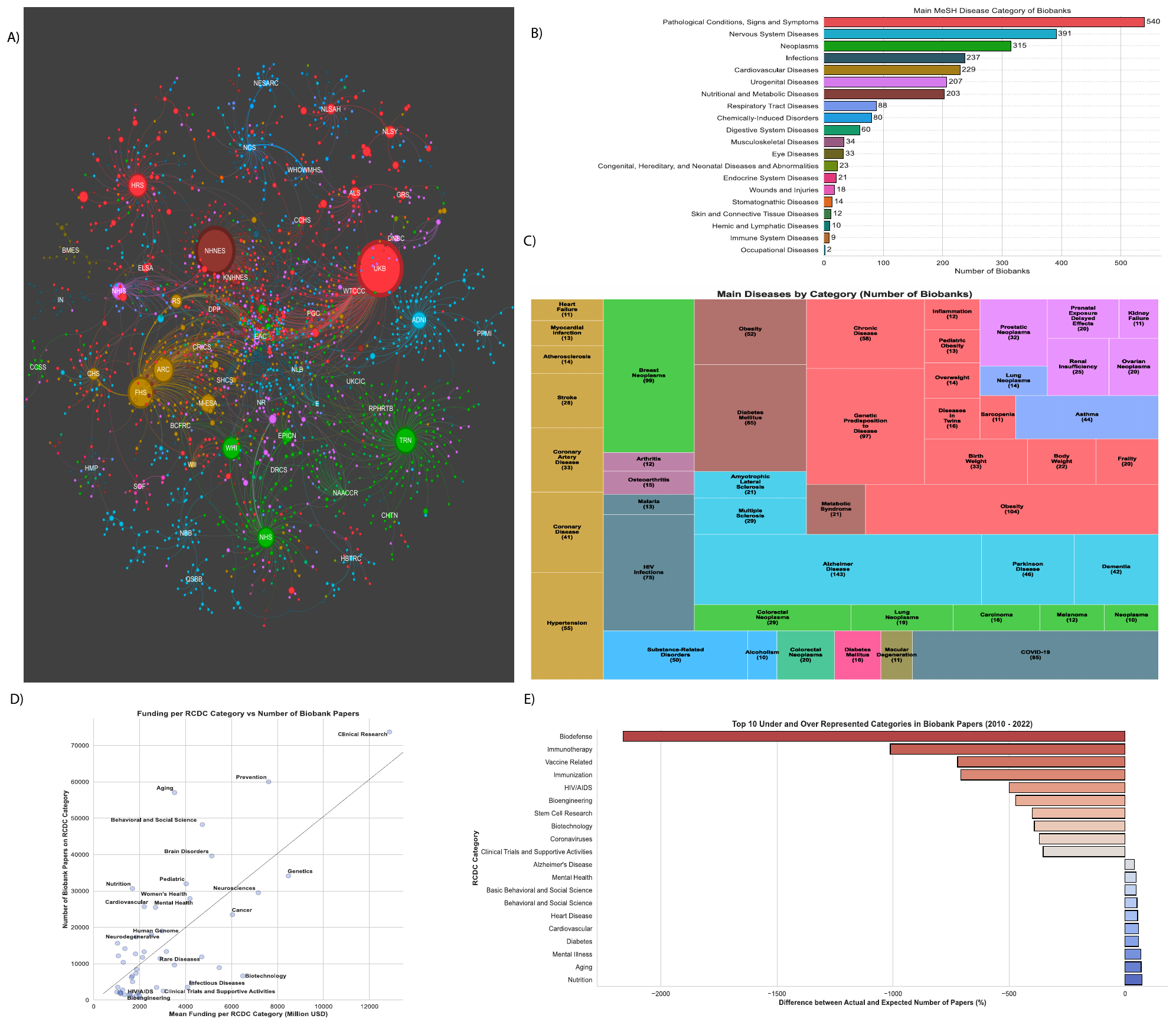}
    \caption{\textbf{The Biobank Disease Universe}. A) The biobank co-citation network whose nodes are biobanks that are connected by an edge if their mentioning publications are frequently cited by the same articles. The size of each node is proportional to its number of mentions, each biobank is colored by its principal disease category (SI~\ref{disease=impact}). B) Number of biobanks by disease category, the colors are matched with the communities in the co-citation network. C) The main conditions studied by biobanks in each disease category, based on the Medical Subject Headers (MeSH) of the articles citing the biobank. We extracted the Research, Condition, and Disease Categorization (RCDC) classification of biobank publications, along with each RCDC category's average annual funding by the NIH, to study:  D) The relationship between the number of biobank publications and funding per RCDC category. E) Over- and under-represented RCDC categories in biobank publications based on their distance to a linear regression's fit.}
    
    \label{fig:diseases}
\end{figure}

The most studied diseases by each biobank reflect not only their focus but also the research interests of the scientific community using them. To capture the impact areas of biobanks, we constructed a co-citation network, whose nodes represent individual biobanks, and connections between nodes occur when biobanks' corresponding publications are cited together (Figure~\ref{fig:diseases}A, SI~\ref{SI:network}). Additionally, we identified the diseases studied by each biobank by analyzing the medical subject headings (MeSH) from publications mentioning these resources (SI~\ref{SI:mesh-diseases}). From this analysis, we identified 2,910 unique conditions across 20 disease categories based on 111,539 research publications. The network is visibly modular, with each community characterized by the focal disease category of its biobanks, as extracted from the MeSH classifications.

\begin{figure}
    \centering
    \includegraphics[width=0.7\linewidth]{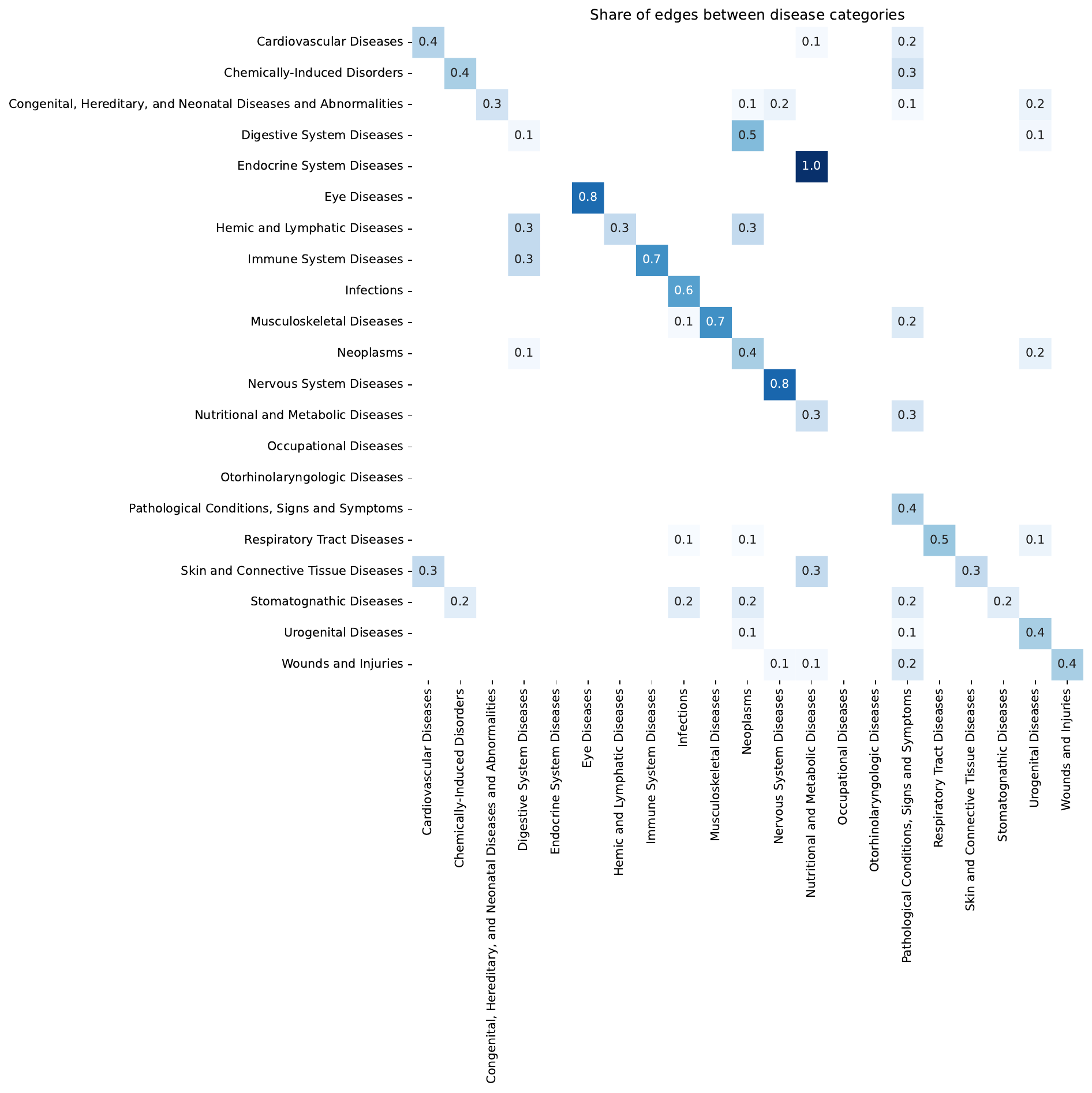}
    \caption{Share of edges between node categories.}
    \label{si-fig:edge-share}
\end{figure}

The disease category ``Pathological Conditions, Signs, and Symptoms'', which covers a broad range of anatomical or physiological states, is the most represented in our network, accounting for 21.3\% of biobanks (540 biobanks). This prominent category is composed of general-purpose biobanks like the UK Biobank. The next largest groups are ``Nervous System Diseases'' (391 biobanks, 15.4\%), ``Neoplasms'' (cancer-related biobanks, 315, 12.5\%), and ``Infections'' (237, 9.3\%). Biobanks focusing on specialized categories like infections or nervous system diseases typically connect with others within the same category, indicating a high degree of specialization (SI Figure~\ref{si-fig:edge-share}). In contrast, general-purpose and cancer biobanks often act as bridges across different biobank communities, exemplified by the nodes corresponding to The Cancer Genome Atlas and the Nurses Health Study, together having 307 connections.

In general, our analysis shows that a high number of biobanks belong to a few categories. Specifically, seven out of ten biobanks are classified as either general-purpose, cancer, nervous system, infections, or cardiovascular disease biobanks. Combined, these categories account for 85\% of all mentions in disease-related studies and target 58\% of all conditions researched. Within these communities, we observe a strong focus on a few conditions---namely obesity, Alzheimer's disease, breast cancer, and diabetes---together being the primary research interests of 20\% of biobank-related publications (Figure~\ref{fig:diseases}C). Despite this high concentration, biobanks demonstrate flexibility in responding to emerging research needs, captured by the rapid attention focus on COVID-19, displayed by infectious disease and respiratory tract biobanks.

\subsection*{Alignment of Funding and Research in Biobank Studies}

To analyze disease representation in biobank research, we extracted the NIH Research, Condition, and Disease Categorization (RCDC) codes of 228,984 publications mentioning biobanks. The RCDC compiles the NIH funding amount for each of the 315 disease categories included from 2010 to 2022. For each RCDC code, we identified its number of biobank-related publications during this time period along with its yearly funding. We limited our analysis to 40 categories, whose average yearly funding exceeds 1 billion US dollars (USD) and 100 publications.

Our analysis revealed a strong linear correlation between the NIH funding of RCDC categories and the number of related biobank articles ($r=0.71$, $p < 10^{-8}$, Figure~\ref{fig:diseases}D), suggesting that higher funding levels generally lead to more research output using biobanks. In particular, clinical research, as the top-funded category with an annual average of 12.8 billion USD, also leads in the number of biobank studies with 73,715 papers. Women’s and minority health categories show similar patterns, ranked 12th and 20th in funding, and 10th and 25th in biobank publications, respectively. Conversely, aging is significantly over-represented in biobank research; it ranks third in publication volume with 57,030 papers, yet only 15th in funding at 3.5 billion USD.

To evaluate the disparity between expected and actual research outputs in biobank studies based on NIH funding for RCDC categories, we applied a linear fit to the funding data and calculated the residuals for each category (Figure~\ref{fig:diseases}E). A positive residual indicates that the actual number of papers exceeded the expected number based on funding, whereas a negative residual indicates fewer papers than expected. Notably, nutrition-related research shows the highest over-representation, with 30,673 actual papers versus 8,209 expected, an increase of 73\%. Other significantly over-represented areas include aging with an increase of 69\%, mental illness at 68\%, diabetes at 58\%, and cardiovascular disease at 57\%. Across 20 RCDC categories with positive residual values, the mean over-representation is +41\%, translating to an average surplus of 11,599 publications per category.

The extent of under-representation in biobank papers across 32 RCDC categories is considerable, with an average shortfall of 7,249 publications per category. More strikingly, the average percentage shortfall is -296\%, indicating that these categories produce approximately three times fewer papers than expected based on NIH funding. This significant under-representation may be anticipated in areas such as biodefense, where the alignment with NIH funding priorities is less direct (-2163\%), or in emerging fields like coronavirus research where biobank resources may not yet meet the research demand (-368\%). Conversely, other categories seemingly well-suited for biobank research, like rare diseases (-98\%), HIV/AIDS (-499\%), stem cell research (-399\%), and precision medicine (-310\%), also exhibit notable under-representation.

Altogether, these results indicate that the majority of scientific impact in biobank research is concentrated in a small number of diseases and research areas. This trend reflects a historical focus on certain disorders and their genetic makeup~\cite{gates_wealth_2021}. However, the research diversity has seen notable improvements in the last decade, aided significantly by new biobanks~\cite{mills_scientometric_2019}. Our findings suggest that funding aligns well with research output for most diseases, effectively enhancing the representation of underrepresented areas within the biobank community.

\subsection*{Exploring the Multifaceted Impact of Biobanks}


While the standard measure of impact is citation based, scientific impact is multifaceted, hence cannot be fully captured by citation-based metrics alone~\cite{weis_learning_2021,aksnes_citations_2019}. This is especially true for biobanks, which often lack standardized citation credit~\cite{cambon-thomsen_assessing_2003,bravo_developing_2015,mabile_quantifying_2013,vora_impacts_2015}. To assess the broader impact of biobanks, we analyzed metadata across 271,394 documents mentioning them. The research impact of biobanks is evident in the 228,632 publications directly mentioning them, which collectively received 10.3 million citations. An additional 194,872 publications were funded by grants where biobanks were mentioned, revealing a different dimension of biobank research impact. These 16,204 biobank-related grants allowed researchers to secure \$30.8 billion USD to support research, personnel training, institutional collaboration, and researcher career development, contributing to the career development of scientists.

The indirect impact of biobanks is captured by their reach across grants, patents, clinical trials, and policies, a metric based on the number of documents referencing a publication mentioning a biobank. For example, a biobank's policy reach equals the number of public policies referencing one of the articles impacted by the biobank, representing a second-degree measure of policy impact. Reach is a good predictor of future impact~\cite{weis_learning_2021}, potentially indicating biobanks' future impact in each sector.

\begin{figure}
    \centering
    \includegraphics[width=1\linewidth]{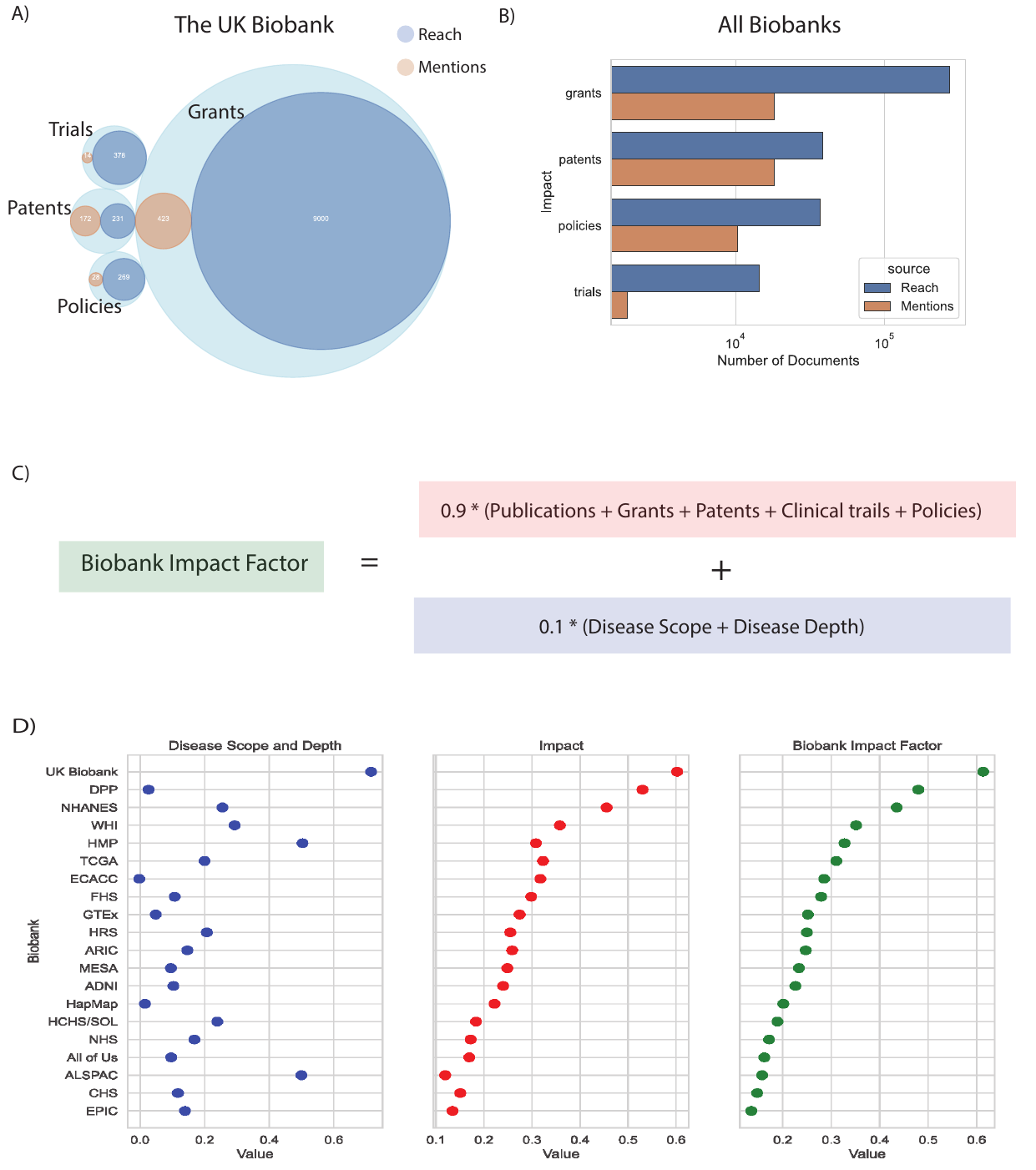}
    \caption{\textbf{Dimensions of the Biobank Impact Factor (BIF)}. We built a Biobank Impact Factor based on the number of mentions a biobank has across science, innovation, public policy, and the depth and scope of its disease impact, including rare diseases. The number of documents directly impacted (mentions) and indirectly impacted (reach) by A) the UK Biobank and B) all biobanks. C) The formula for obtaining the BIF based on impact, disease scope, and depth. D) The BIF of the top 20 biobanks, impact, disease depth \& scope values.}
    \label{fig:top}
    
\end{figure}

The indirect impact captured by reach is illustrated for the UK Biobank (Figure~\ref{fig:top}A). The UK Biobank has been directly mentioned in 423 grants (direct impact), a number eclipsed by the 9000 grants that cite one of its related publications (indirect impact). Similarly, only 14 clinical trials refer to the UK Biobank directly in their summary (direct impact), but 378 cite one of its publications (indirect impact), a 26-fold increase in impacted trials. On average, the indirect impact (reach) of the UK Biobank is 14.8 times larger than its direct impact (mentions), showcasing its broad influence across multiple health-related sectors.

More generally, we find reach to be significant across all biobanks, revealing that at least one biobank-related publication is referenced in 72,817 grants, 23,576 patents, 16,752 policies, and 10,679 clinical trials (Figure~\ref{fig:top}B). The reach of biobanks is considerably larger than their direct impact: 5.5 times larger for patents, 14.2 for policy documents, 6.6 for clinical trials, and 33 in the case of grants.
These results confirm the multidimensional aspect of biobank impact, supporting the claim that a comprehensive metric of biobank impact should also include health sectors outside of academic research.



\subsection*{Biobank Impact Factor}

Biobank mentions across various document types reveal their multidimensional impact. Strong correlations exist between mentions in publications and grants (Pearson correlation $r = 0.8$) and between publications and clinical trials ($r=0.5$), highlighting the crucial role of biobanks in research. However, correlations between publications and public policy ($r=0.1$) and publications and patents ($r=0.4$) are weaker, suggesting that some biobanks significantly impact specific domains despite lower publication impact.

To measure the true impact of a biobank, we introduce the Biobank Impact Factor (BIF, Figure~\ref{fig:top}C), capturing biobank references in scientific research, innovation, and public health. To minimize bias towards large biobanks and cohort studies, we include a metric of disease impact based on existing indicators of high-value bioresources~\cite{mabile_quantifying_2013,rush_improving_2020}. Specifically, we account for the depth and scope of disease research that relies on a biobank, including its use to advance knowledge of rare diseases (SI~\ref{SI:disease-impact}). BIF is a weighted average of a biobank's impact across publications, grants, patents, clinical trials, public policy, and disease, normalized between -1 and 1 (SI~\ref{SI:BIF}). To minimize bias due to the length of time a biobank has been available, we normalize its BIF based on the year that the biobank was first mentioned. We computed the BIF for all 1,326 biobanks in our dataset with at least 20 publications, a cutoff chosen to have enough disease data while retaining at least half of the biobanks.

Among the biobanks with the highest BIF (Figure~\ref{fig:top}D), we find the Diabetes Prevention Program (2nd, BIF $=0.48$), the Women's Health Initiative (4th, BIF $=0.35$), the Human Microbiome Project (5th, BIF $=0.32$), the Cancer Genome Atlas Program (6th, BIF $=0.31$), the Framingham Heart Study (8th, BIF $=0.28$), and the Genotype-Tissue Expression Project (9th, BIF $=0.25$), all supported by the National Institutes of Health. The list also includes two UK-based biobanks: The UK Biobank (1st, BIF $=0.61$) and the European Collection of Authenticated Cell Cultures (7th, BIF $=0.29$), as well as two other US-based studies, the National Health and Nutrition Examination Survey (3rd, BIF $=0.44$), and the Health and Retirement Study (10th, BIF $=0.25$), completing the top-10 list. To allow easy access to the collected data and metrics, we developed an online tool to search, explore, and compare the scientific impact of all biobanks, distinguish the biobank communities and diseases studied, together with their measures of scientific impact, available as a dashboard at \url{http://biobanks.pythonanywhere.com/}.

 \subsection*{Biobank Impact is National and Institutional}

A current survey on biobank use concluded that researchers have a strong preference for local and familiar sources~\cite{lawrence_barriers_2020}, prompting us to measure the extent to which biobanks have local versus global impact. We first identify the host institution of each biobank (SI~\ref{SI:biobank-teams}) and extract the share of scientific impact, measured by mentions across publications, coming from the host institution or the host country. We find that on average 73.5\% of the publication impact comes from researchers in the host country of the biobank, and 29.4\% have the same institutional affiliation (Figure~\ref{fig:local-hidden}~A). We compare these results to a null model where we randomly rewire the citation network while preserving the number of citations of each biobank, finding that the local impact by country and affiliation are statistically highly significant ($p$-value $< 10^{-10}$, SI~\ref{SI:null-model}). 

To compare the impact levels of biobanks, we used the BIF of each biobank, allowing us to assess the share of mentions coming from the same country and affiliation for top and bottom biobanks (Figure~\ref{fig:local-hidden}~B). The percentage of same-country mentions of bottom-20\% biobanks, their national impact, is on average $76.8\% \pm 3.2\%$, a value higher than the top-20\% biobanks ($71.1\% \pm 3.1\%$). Even if the impact is mostly national for a biobank, the international impact of bottom biobanks is 6\% weaker than that of top biobanks ($p<0.02$). A finer assessment of local impact is the share of institutional impact, i.e., the share of mentions coming from research teams from the same institution of the biobank itself. We find that the average percentage of the institutional impact of bottom-20\% biobanks is $37\% \pm 27\%$, significantly larger than top-20\% biobanks ($20\% \pm 19\%$, $p<10^{-16}$), indicating that the impact of top biobanks is at the same time less institutional and more international.

\subsection*{Access to Biobanks Driven by Co-authorship}

Biobanks often restrict scientists' access to their data, partly driven by privacy considerations (personal data security is one of the main concerns for controlling access), and less justifiably so, because maintaining and supplying the data is costly~\cite{bravo_developing_2015}. Yet, the often lengthy application process to obtain access to the data is often bypassed via co-authorship with the biobank team. This practice has a profound effect on the authorship of the 147,656 articles mentioning a biobank for which we identified its supporting team (SI~\ref{SI:biobank-teams}). Indeed, we find that, on average, at least one member of the biobank is a co-author on 59.6\% of the articles mentioning the biobank, usually being the PIs of the biobank (Figure~\ref{fig:local-hidden}~C). This is congruent with survey data, concluding that co-authorship is a prime incentive for data sharing from biobanks~\cite{kleiderman_author_2018}.

On average, members from biobanks with top-20\% biobank BIF are listed as co-authors in 44\% of the papers mentioning the biobank, compared to 67\% papers for bottom-20\% biobanks (Figure~\ref{fig:local-hidden}~D). This suggests that a lower share of co-authorship of a biobank may be indicative of the demand for the data. On average, non-PI members of top biobanks produce 24 external collaborations without the biobank PIs, a significant percentage (24\%, $p < 0.036$) compared to lower impact biobanks (12\%, $p=0.06$), where non-PIs collaborate in only 3 papers without the biobank PIs.
A significant majority of all the recorded collaborations take place within the same country (39,496 out of 49,192, or 80\%), which may explain the strong national impact of the bottom-20\% biobanks, whose impact is predominantly collaborative. In contrast, only 185 (0.3\%) of all collaborations happen with other PIs from the same institution, suggesting that most institutional impact comes from the biobank team itself. In other words, collaborative work increases the recognition and scientific impact of the biobank creators, with the downside of setting geographical barriers to their use.

\subsection*{Citations Underestimate the True Scientific Impact of Biobanks}

The reference paper of BioBank Japan~\cite{nagai_overview_2017} has gathered $470$ citations and the article introducing the UK Biobank~\cite{sudlow_uk_2015} has been cited $3,052$ times, raising the question, does citation-count capture the true scientific impact of biobanks? We find that not all papers that use biobank data cite the reference papers published by the biobank. Therefore, to estimate the true scientific impact of biobanks, we measured hidden citations, which are articles that mention the biobank in their title, abstract, or acknowledgments sections without citing the resource article of the biobank.

\begin{figure}
    \centering
    \includegraphics[width=1\linewidth]{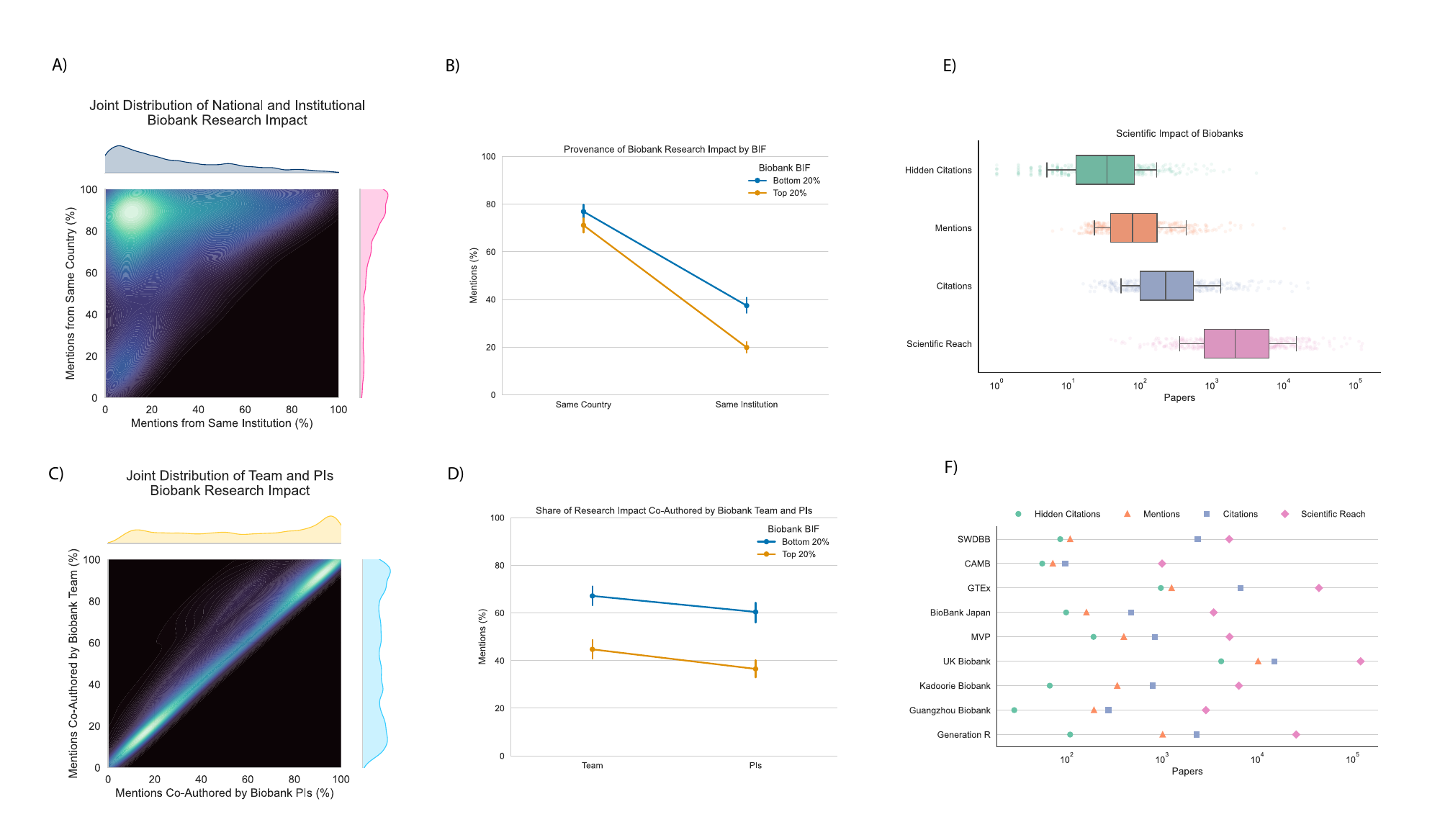}
    \caption{\textbf{Provenance of biobank research impact and hidden citations}. A) The joint distribution of national (same country, purple) and institutional (same research affiliation, pink) impact of biobanks based on mentioning papers.
    B) Mean percentage and 95\% confidence intervals of mentions coming from papers in the same country and institution for bottom-20\% (orange) and top-20\% biobanks (blue) based on biobank impact factor. Error bars represent 95\% confidence intervals. C) Joint distribution of the percentage of mentioning papers listing at least one principal investigator (PI, light yellow) or a team member (light blue) of the biobank. D) Mean percentage and 95\% confidence intervals of mentions listing a PI or a biobank team member for bottom-20\% (orange) and top-20\% (blue) biobanks.
    E) Distribution of papers mentioning a biobank but not citing its reference papers (green), mentioning a biobank (orange), citing its reference papers (purple), or citing its mentioning papers (biobank reach, pink). Presumably, mentioning articles should include a reference to one of the reference papers of each biobank so the number of hidden citations should be small, as they account for papers for which the biobank is central but fail to cite its main articles.}
    \label{fig:local-hidden}
\end{figure}

We find 14,995 articles that cite at least one of the 5 UK biobank reference papers~\cite{sudlow_uk_2015,bycroft_uk_2018,fry_comparison_2017,littlejohns_uk_2020,petersen_imaging_2013}. Yet, we find an additional 10,123 articles that mention the UK Biobank after the first reference paper was published. Most importantly, only 59\% of those articles cite one of the biobank's reference papers, indicating that 41\% of the users do not acknowledge through the traditional method of impact, i.e.\ citations, their reliance on the biobank. To assess the hidden impact of other biobanks, we identified the 962 reference papers of 500 biobanks and computed their number of hidden citations across 96,745 mentioning articles (Figure~\ref{fig:local-hidden}E and SI~\ref{SI:hidden-citations}). We find that, on average, 41.2\% of the 203 articles mentioning a biobank do not cite any of their reference papers, suggesting a general pattern of undercitation. Some of the most undercited biobanks include the GTEx Project (Figure~\ref{fig:local-hidden}F, 967 hidden citations, 77.1\% of users), the South West Dementia Brain Bank (SWDBB, 85 hidden citations, 78.7\%), and the Copenhagen Aging and Midlife Biobank (CAMB, 55 hidden citations, 77.4\% of users). Biobanks with a lower number of hidden citations include the Guangzhou Biobank (28, 14.5\%), the China Kadoorie Biobank (66, 19.5\%), and the Generation R Study (108, 10.6\%), indicating that, traditional measures of impact, i.e.\ citations, highly underestimate the true academic impact of biobanks.

To obtain a more accurate measure of the scientific reach of biobanks, we considered all articles either citing their reference papers or the papers mentioning biobanks, obtaining 120,551 unique papers for the UK Biobank and 3,465 for BioBank Japan, representing more than 7 times their current detectable citations. On average, we find that the scientific reach of a biobank is 13 times greater than the number of citations (Fig~\ref{fig:local-hidden}). Note that these numbers still underestimate the true impact of these biobanks as we have not scanned the full text of the scientific papers, which may reveal additional papers that take advantage of biobank resources without citing them.


\subsection*{Biobank Features and their Relation to BIF}

The reach and the scientific impact of a biobank are driven by multiple variables, ranging from cohort size (number of individuals sampled) to type (health-based or population-based), genetic data availability, and the biobank's PI level of recognition. Understanding which of these variables plays a more defining role can help biobank creators identify and implement the best strategies to increase their reach and scientific impact. To differentiate the role of those variables, we built a biobank Inherent Characteristics Model (ICM), based on a generalized linear model with a Gaussian family designed to explain the BIF of a biobank ($Y$) given its set of characteristics,

\begin{equation}\label{eq:model1}
\log(Y + 1) \propto \beta_0 + \beta_1 \times X_1 + \beta_2 \times X_2 + \dots + \beta_{14} \times X_{14} + \epsilon.
\end{equation}

On the r.h.s\ we list the 14 features of biobanks that could affect their impact (SI~\ref{SI:biobank-data}), namely large cohort sample (sample size in the top 10\%), open data index (2 levels, low, and high), the biobank's PIs citation level at the time it was created (2 levels, low and high), population-based (1 for population-based biobanks, 0 for health-system based), available genetic data (sub-divided in GWAS, Whole-genome sequencing, and gene-environment data), registries, surveys \& questionnaires, follow-up data, and medical records. The error term $\epsilon$ follows a standard normal distribution. 

\begin{figure}[h!]
    \centering
    \includegraphics[width=1\linewidth]{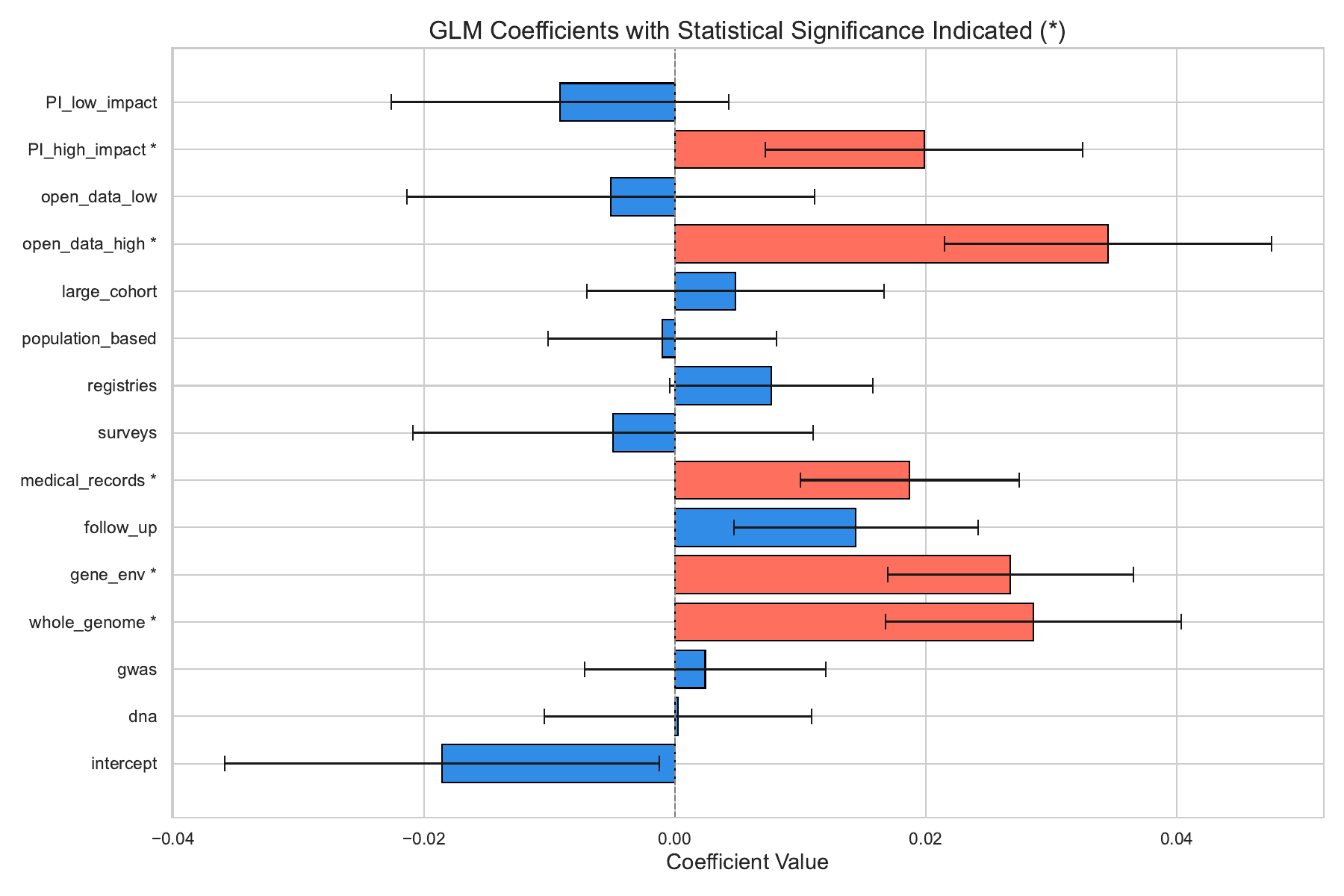}
    \caption{\textbf{Related biobank features to biobank impact factor}. We present a generalized linear model to identify the key features explaining biobank impact. The coefficients of different features are captured by model~\ref{eq:model1} considering different binary characteristics of biobanks, including whether the cohort size is large (top 10\%), sampled from a general population, data access is open to external researchers, the average citation count of the biobank PIs (bottom and top-10\% with respect to all PIs), along with availability of genetic data (subdivided into genetic markers or DNA, GWAS, whole-genome sequencing, or gene-environment interactions), follow-up data, disease-specific data (based on registries), surveys \& questionnaires, and linked medical records. Each feature's coefficient is shown together with its 95\% confidence interval. Significant features after applying Bonferroni correction are indicated with a star symbol and red color.}
    \label{fig:model}
\end{figure}

The model was fitted using data from 468 biobanks ($R^2=0.41$), and the $p$-values of the coefficients were Bonferroni corrected. The deviance of the model was $0.686$, which, along with a Pearson Chi-square of $0.687$, suggests a good fit to the data (SI~\ref{glm}). We find five statistically significant coefficients, capturing the more important characteristics related to BIF (Figure~\ref{fig:model}). The largest model coefficient is the one related to a high open-data index ($\beta_{\text{oa}} = 0.0345$, $p = 0.002$), followed by whole genome sequencing data ($\beta_{wg} =0.0286$, $p=2\times 10^{-6}$). It is important to note that BIF is a normalized metric of impact with a maximum of $0.47$ for the UK Biobank, so possessing a feature with a coefficient of $0.01$ sets a biobank in the top half of the BIF distribution.

The open data index is proportional to the number of external requests for the biobank data, and its significant correlation with BIF demonstrates the importance of data applications not necessarily leading to co-authorship (SI~\ref{si:open-data-index}). In other words, well-defined data policies, especially those open to new researchers, can lead to a higher impact with the downside of renouncing direct recognition to the biobank's team through collaborations.

Relatedly, we find that having a highly cited founder is significantly correlated with an increase in BIF ($\beta_{pi}=0.0198$, $p=0.002$). Besides expertise, the strong reputation of the biobank's PIs provides an extensive network of collaborations, along with trustful credentials for potential users, a factor shown by previous surveys to be a major reason for data application~\cite{lawrence_barriers_2020}. On the other hand, we find that having weakly cited PIs at the time of biobank formation is not significant to BIF ($p=0.18$).

The principal feature of a biobank, however, is the type of data it provides. Besides whole genome sequencing, gene-environment interaction data, joining genetic and environmental information, is a type of genetic data strongly associated with biobank impact ($\beta_{ge} = 0.0267$, $p=9 \times 10^{-8}$). Data from genetic markers alone is not significantly correlated to BIF ($p=0.96$), perhaps due to the fact that most biobanks in our list ($74\%$) include genetic markers. Similarly, GWAS data is not significantly related to BIF ($p = 0.62$) and is included in most biobanks ($52\%$).

The last important feature related to BIF is the inclusion of the medical records of the cohort sample ($\beta_{mr}=0.0187$, $p=2 \times 10^{-5}$), providing evidence to survey-based findings listing the availability of medical records as the most cited reason to apply for biobank data~\cite{lawrence_barriers_2020}. We find the other data features, to not be significantly related to BIF, including follow-up data ($p = 0.0037$, not significant after Bonferroni correction), surveys ($p=0.5$), registries ($p=0.06$), and a large cohort sample ($p=0.426$), suggesting that neither data types are a condition to biobank impact.

\section*{Conclusions and Discussions}

A key asset of a successful biobank is the research community it engages. The finding that high-impact biobanks have strong ties to their founders' scientific community, reflected by the high co-authorship rates of the articles citing the biobank, suggests that a biobank's long-term impact relies strongly on the scientific ties of their founding and operating teams to the core group of scientists who benefit from the collected data. For a new biobank to succeed, it is therefore paramount for its founding team to be well integrated within the biomedical community it serves, best achieved by the lead of an experienced and reputable scientist. Activating the collaboration network of the founders appears to not only be essential to promote the biobank but also to extract its potential scientific value by attracting external collaborators.

The exceptional number of biobanks, and the limited visibility of most of these valuable resources, raise the question, of how to increase awareness of the value of the data offered by them. This could be enhanced by a global biobank locator system, that helps identify the best biobank for a particular research project. Although several biobank databases exist, they are limited to particular geographical areas and lack the resource articles with the biobank~\cite{wichmann_comprehensive_2011,odonoghue_how_2021}.

To address this need, we are releasing as an online tool the biobank dataset we collected together with all the impact metrics included here, allowing researchers to assess the diseases studied by a biobank community, find adjacent resources, and explore the scientific impact of each biobank. The application could allow biobanks attract potential users and find collaborative opportunities by identifying research teams that belong to the same scientific community. A list of features includes a summary of the origin and usage of biobanks, their collaborators, and team characteristics. A ranking based on the biobank BIF is available, together with different networks representing biobanks under different connectivity frameworks, e.g., based on co-mentions in patents, clinical trials, or grants. 

High-throughput genetic data, coupled with statistical tools and increased computing power, have dramatically increased the size and power of genetic-based population studies~\cite{lawson_is_2020, pingault_using_2018}. Not surprisingly, our work shows that whole genome sequencing and gene-environment interaction data are an important element of high-impact biobanks. Medical records data that could be used to explore multiple diseases is another key characteristic. Mega-biobanks with a large corpus of genetic data covering a wide diversity of diseases are expected to play a significant role in the future of biobanks. These biobanks, like the Million Veteran Program (MVP) in the United States~\cite{gaziano_million_2016}, can store genetic information from hundreds of thousands or even millions of individuals. Having an open approach to data sharing is essential to reach external researchers, as our results show. Making their data more accessible to independent studies gives biobanks the potential to help unlock new insights into the causes of diseases and pave the way for more effective treatments and personalized medicine.

Indeed, our results indicate that a general-purpose biobank with genetic data is an optimal resource for high-impact population studies. Note, however, that scientific impact is not the only measure of success. For instance, a biobank created to target a rare disease may not be seen as a high-impact biobank by the traditional measures of academic impact but, in reality, could pave the way towards finding critical cures, having a profound impact on both participants and researchers. For this reason, it is important to include these impact metrics to assess biobank impact. This motivated the proposed metric of impact, the BIF, as a measure to capture the depth and scope of disease research produced with biobanks, as well as the impact across areas like innovation and public policy.

Although our analysis of the factors related to biobank impact considers multiple data characteristics, it lacks information on processing times, data quality, and access costs, important factors for potential users~\cite{lawrence_barriers_2020}. As biobanks become a central resource for biomedicine and their data policies become more widespread, these factors should be included in future assessments of biobank impact.

\subsection*{Acknowledgments}

We thank the wonderful research community at the Center for Complex Network Research, particularly those in the success and biology groups, for valuable discussions and comments. We would also like to thank the community of biobank developers and maintainers, without which this and many other studies would not be possible.

\subsection*{Funding}
This work was funded by the United States Department of Veteran Affairs.

\newpage

{\color{blue}
\begin{table}[ht]
\centering
\caption{Collected Features of Biobanks}
\label{tab:biobank_features}
\begin{tabular}{>{\raggedright\arraybackslash}p{3cm} >{\raggedright\arraybackslash}p{8cm}}
\toprule
\textbf{Type} & \textbf{Features} \\
\midrule
Cohort & Size, age group, origin, type (population/health-based) \\
Data & Genetic (Whole-genome sequencing, GWAS, or environmental), socioeconomic, follow-up, medical records, surveys, registries \\
Papers & MeSH identifiers, RCDC categories, HRCS classification \\
Patents & Cooperative Patent Classification codes, assignee \\
Grants & NIH activity codes, USD amount, funders \\
Clinical Trials & Study type, RCDC categories  \\
Public Policy & Organization type, country \\
\bottomrule
\end{tabular}
\end{table}
}

\begin{table}[ht]
    \centering
    \caption{Overview of grants mentioning biobanks}
    \begin{tabular}{lrr}
    \toprule
     & Total & Biobank Mean (STD)  \\
    \midrule
        Grants & 16204  & 14.3 (48.5) \\
        Million USD & 30,863 & 28.3 (115) \\
        Funding Organizations & 298 & 3.4 (5.6) \\
        Funding Countries & 37 & 1.6 (1.5) \\
        Research Organizations & 2,421 & 8.7 (21.3) \\
        Research Countries & 74 & 2 (2.7) \\
        Training grants & 1,359  & 1.3 (6.7) \\
        Collaboration grants & 1106 &  1 (4.6)\\
        Resulting publications & 193,884 & 238.6 (894.4) \\
    \bottomrule
    \end{tabular}
    \label{table:grants}
\end{table}

\clearpage
\newpage

\printbibliography

\end{refsection}

\clearpage

\begin{refsection}
{\huge \textbf{Supplemental Information}}\\

{\LARGE Quantifying the Impact of Biobanks and Cohort Studies}\\

The code presented here can be found at \url{https://github.com/Barabasi-Lab/quantifying_biobanks}. The code and the accompanying data is also stored in a Zenodo repository at \url{10.5281/zenodo.11671293}. A dashboard is available at \url{http://biobanks.pythonanywhere.com/}.

\section{Biobank dataset}\label{SI:biobank_dataset}
The process to build a corpus of biobank names was based on two steps. First, we compiled a set of 16 biobank catalogs (SI~\ref{SI:catalogs}), and second, we extracted biobank names from the text of 3,310,320 biobank-related articles (SI~\ref{SI:expansion}). After each step, we processed the names identified to find and remove duplicates based on word similarity and co-appearance in articles of each pair of names (SI~\ref{SI:deduplication}). The resulting corpus of biobanks contains 2,663 unique biobank names that were then used to gather biobank mentions from textual documents (SI~\ref{SI:mentions-data}). Finally, the dataset was validated using an external database of GWAS studies (SI~\ref{SI:data-validation}).

\subsection{Biobank Catalogs}\label{SI:catalogs}

Although there is yet no central biobank repository including all biobanks, there have been several efforts to build catalogs for the scientific community. The largest to date is the BBMRI-ERIC Directory of Biobanks and their collections. The resource was established to ``improve accessibility and interoperability between academic and industrial parties to benefit personalized medicine''~\cite{van_ommen_bbmri-eric_2015}. As of 10/2023, the directory included a list of 617 European biobanks together with their 3,176 collections and related metadata, providing a large set of biobank names.

To complement this corpus, we integrated the biobank names found in 15 prominent directories of population bioresources (the resulting corpus of biobanks can be found in raw and processed formats in the data folders `cohorts' and `raw\_cohorts', respectively). Specifically, this dataset includes biobanks from BIOLINCC~\cite{giffen_providing_2017}, EPND~\cite{bose_data_2022}, IADRP~\cite{liggins_international_2014}, Molgenis~\cite{van_der_velde_molgenis_2019}, P$^3$G (Public Population Project in Genomics and Society)~\cite{ouellette_p3g_2014}, DCEG~\cite{wang_improved_2012}, DPUK~\cite{bauermeister_dementias_2020}, UKRI cohort directory, Birthcohorts~\cite{vrijheid_european_2012}, CEDCD~\cite{kennedy_cancer_2016}, The Pooling Project of Prospective Studies of Diet and Cancer~\cite{smith-warner_methods_2006}, Wikipedia's list of biobanks, SciCrunch~\cite{bandrowski_decade_2022}, Maelstrom~\cite{bergeron_fostering_2018}, dbGAP~\cite{tryka_ncbis_2014}, and JPND~\cite{lerche_methods_2015}. The resulting list contained 1,576 unique names. Most of these resources were available for download with the exception of  Birthcohorts, IADRP, Pooling Project of Prospective Studies of Diet and Cancer, and UKRI for which we crawled the websites (code found at \url{python/crawl}).

Furthermore, we extracted the biobank names found in the titles of papers published in the `Cohort Profile series' of the International Journal of Epidemiology, and BMJ Open Journal, both having strict guidelines to start the title with `Cohort Profile:', followed by the biobank name. We identified these publications using regular expression on the full publication dataset of Dimensions (files \url{SQL/cohort_profile.sql} and \url{SQL/cohort_profile2.sql}), resulting in 488 publications containing 368 biobank names.  

Finally, we used ChatGPT to generate a list of biobanks using the prompt `produce a list of biobank names along with their country of origin', resulting in 738 names. After processing to remove special characters, trailing spaces, and parenthesis, the resulting list contained 2,207 unique biobank names (file \url{data/cohorts/final\_cohorts.csv}). The code to produce the list is found at \url{python/cohort\_names/raw\_cohorts.py}.

\subsection{Biobank names expansion}\label{SI:expansion}

In order to expand our biobank list, we did a bibliography search based on a set of 3,310,320 biobank-related publications. These publications were obtained by first identifying 162,988 articles mentioning one of the 2,207 biobanks previously obtained, and then extracting their citing articles. The resulting set of publications is then either mentioning directly a biobank or citing a paper mentioning one. The corpus of mentioning publications was obtained by searching in the title, abstract, and acknowledgments sections of articles in Dimensions (\url{SQL/mentions/publications.sql}). The code to obtain the set of biobank-related publications is \url{SQL/expansion/papers_citations.sql}.

We used regular expressions based on common keywords found in the names of biobanks and similar repositories, including Biobank, Tissue Bank, Registry, Biorepository, Project, and 12 other keywords. The complete list of regular expressions as well as the code to obtain the potential biobank names can be found at \url{SQL/expansion/clean_expanded_cohorts.sql}. The resulting list of names yielded 23,435 potential biobank names. We filtered potential biobanks in this corpus of names based on each biobank's number of mentions in two different corpora: the set of biobank-related papers (N=3,310,320) and the set of papers mentioning a biobank from the catalogs (N=162,988). A potential biobank was removed from consideration if less than 5\% of the articles mentioning it contained a mention to a biobank from the catalogs and less than 50\% of its mentions came from biobank-related publications. The code to obtain papers mentioning the potential biobanks is in  \url{SQL/expansion/expanded_papers.sql} (set of papers mentioning a potential biobank) and \url{SQL/expansion/expanded_papers_internal.sql} (set of mentions from biobank-related papers). The code to obtain their number of co-mentions with biobanks from the catalogs is at \url{SQL/expansion/clean_expanded_cohorts.sql} and the code to calculate the percentages of mentions on each corpus is \url{SQL/expansion/cohorts_counts.sql}.

This expansion process resulted in an additional 1,924 biobank names (file \url{data/expansion/selected_cohorts.csv}), resulting in a total of 4,131 biobanks. The code to clean, preprocess, and obtain this list of biobanks is found at \url{python/expansions/total_inner_ratio.py}.

\section{Biobank mentions}\label{SI:biobank_mentions}

In order to assess the impact of biobanks we searched their mentions across multiple textual documents, from the Dimensions academic database we include scientific publications, clinical trials, grants, and public policy documents. From Google Patents Public Data we searched for patents mentioning biobanks, as they provide the full-text of each patent. The search for mentions was case insensitive and included the name of the biobank preceded by the word `the' to reduce false positives. The code to search for mentions can be found in multiple files named after the type of document they were searched on in the folder \url{SQL/mentions/} for 2,207 catalog biobanks, and \url{SQL/expansion/mentions/} for the expanded list of 1,924 biobanks.

To focus on biobanks with traceable presence in biomedical science, we remove those without at least one mention in either the title, abstract, or acknowledgments section of a scientific publication. This included 1,163 biobanks from the catalogs and 1,882 biobanks from the expanded list, for a total of 3,045 biobanks.

\subsection{Removing Biobank Duplicates}\label{SI:deduplication}

A common issue with biobanks is the multiplicity of names a single biobank can accumulate across publications. The reasons behind this issue are multiple: from  name alterations in follow-up studies (e.g., Framingham Heart Study, Framingham Children's Study, and Framingham Offspring Study), use of abbreviations (e.g., ARIC instead of Atherosclerosis Risk in Communities), to inconsistencies in their naming across publications (e.g., Northern Sweden Medical Biobank and Medical Biobank of Northern Sweden or Leeds Biobank and Leeds Multidisciplinary Research Tissue Bank). Additionally, biobanks and their corresponding cohort studies can be used interchangeably, like in Lifelines Biobank and Lifelines Cohort Study, similarly to biobanks that are part of a consortium  (e.g., African Neurobiobank for Precision Stroke Medicine and H3Africa).

In order to minimize the number of name pairs referring to the same biobank, we performed a deduplication process based on two factors: (i) string similarity and (ii) co-mentions. The former allows to identify similar names of a single biobank and the latter to identify duplicates lacking string similarity such as abbreviations from their full name. String similarity of each biobank pair $a$ and $b$ was calculated using the Indel distance $I(a,b)$, measuring the minimal number of insertions and deletions needed to transform one string into the other, normalized by the maximum number of possible deletions and insertions ($I(a,b)=1$ implying that $a = b$). We used two variants of these distance: the partial ratio $I_{pr}$, measuring the Indel distance needed to match the smallest string into a sub-sequence of the larger string (here $I_{pr}(a,b)=1$ does not necessarily imply that $a = b$); and the token set ratio $I_{sr}$, based on the minimal number of word deletions and insertions needed to match $a$ and $b$ (the word order is irrelevant). Using Spacy, a natural language model, we `cleaned' the names prior to their analysis to remove coordinating conjunctions, auxiliaries, punctuation, symbols, and numerals.

Next, we measured the co-mention similarity of each biobank pair across the 228,761 publications and 19,299 patents mentioning biobanks. The co-mention similarity $S(a,b)$ between two biobanks $a$ and $b$ is equal to the proportion of documents mentioning $a$ that also mention $b$ in their text. In other words,  $S(a,b)=1$ implies that $a$ is exclusively mentioned in publications mentioning $b$, and $S(a,b)=0$ that no publication co-mentions $a$ and $b$. Note that $S$ is not symmetric: for cases where mention counts of biobanks $a$ and $b$ differ, we have that $S(a,b) \neq S(b,a)$.

Finally, we identified a pair of biobanks $a$ and $b$ as duplicates if at least one of the following statements occurred: (i) min$\{S(a,b), S(b,a)\} > 0.2$ and $I_{pr}(a,b) > 0.9$, (ii) min$\{S(a,b), S(b,a)\} > 0.05$ and $I_{pr}(a,b) = 1$, (iii) max$\{S(a,b), S(b,a)\} > 0.3$ and $I_{pr}(a,b) > 0.9$, (iv) max$\{S(a,b), S(b,a)\} > 0.1$ and $I_{sr}(a,b) = 1$, where co-mentions where extracted from publications. For patent-based similarity,  we only considered $a$ and $b$ to be duplicates if all three inequalities were true: max$\{S(a,b), S(b,a)\} > 0.4$, min$\{S(a,b), S(b,a)\} > 0.3$, and $I_{pr}(a,b) = 100$, where stricter similarity conditions accounted for a higher probability of co-mentions in full-text documents.

The deduplication process resulted in 337 duplicate pairs, often forming duplicated groups where a biobank had 3 or more alternative names (An example being the Framingham Heart Study or FHS, with 10 variations). To obtain all duplicate groups, we constructed the duplicate network in which nodes represent biobank names and edges duplicated pairs, and each connected component represents name variations of the same biobank. We completed the deduplication process by selecting the biobank name with the greatest number of article mentions in each duplicate group to be the name variation to replace the others. 

After deduplication, the final biobank corpus contained 2,263 names
mentioned across 228,761 articles, 16,210 grants, 15,469
patents, 1,769 clinical trials, and 9,468 public policy documents. The code containing the deduplication process is found at \url{python/cohort_names/deduplicate.py}.

\subsection{Scientific articles}\label{SI:publications}

In order to look for mentions across Dimensions' publications (from October 2023), we first processed the text from the title, abstract, and acknowledgments sections to remove Unicode characters, double spaces, commas, and periods. The code is found at \url{SQL/clean_text/publications.sql}. Once the articles' text is pre-processed, we look for textual mentions of biobanks within the title, abstract, and acknowledgment sections of 141,219,539 articles resulting in 250,857 biobank-article mentions on 228,761 articles (\url{data/expansion/cohort_patents.csv}).

\subsection{Patents}\label{SI:patents}

We use the publicly available Google Patents Open Data to search for biobank mentions on the full text of 153,696,878 patents. The search was done using Google BigQuery on the table \texttt{patents-public-data.patents.publications}, resulting in 18,183 biobank-patent mentions on 15,469 unique patents (\url{data/expansion/cohort_patents.csv}. The patents found were then merged to the patent data from Dimensions using the grant application number id (\url{SQL/join_dim_google_patents.sql}). 

\subsection{Grants}\label{SI:grants}

We used grant data from Dimensions, data composed of 5,040,039 grants. We found 18,251 biobank-grant mentions on 6,210 unique grants (\url{data/expansion/cohort_grants.csv}).

\subsection{Clinical trials}\label{SI:clinical-trials}

We used clinical trials from Dimensions, data composed of 801,708 clinical trials. We retrieved 1,881 biobank-clinical-trial mentions from 1,769 unique clinical trials (\url{data/expansions/cohort_clinical_trials.csv}).

\subsection{Public Policy Documents}\label{SI:policy-documents}

We used the public policy documents from Dimensions, a dataset composed of 1,783,533 public policy documents. We identified 10,285 biobank-public-policy mentions on 9,468 unique public policy documents (\url{data/expansions/cohort_public_policy.csv}).

\section{Data Validation}\label{SI:data-validation}

To assess the comprehensiveness of our database of biobanks and related documents,
we used the NHGRI-EBI Catalog of Genome-Wide Association Studies (GWAS) as a reference to measure the scope of our data (the catalog was accessed June 2024), compiling more than 100,000 genetic associations to disease from 6,899 research publications~\cite{sollis_nhgri-ebi_2023}. Indeed, biobanks and population studies play a major role in the discovery of biomarkers by providing samples and genetic data used in GWAS. They are not the only contributors, however, cross-sectional and case-control studies---largely found in clinical trials and not really considered in our dataset---remain the most used resources for biomarker discovery~\cite{wijmenga_importance_2018}. Another potential caveat to consider is our lack of full-text publications for our search, a limitation not present in the reference GWAS catalog, directly receiving its data from the authors of the articles.

To be able to compare the overlap between the two datasets, we used Dimensions to search the PubMed Ids of our 228,761 biobank-related publications, successfully doing so for 76.6\%, but dropping 1/4 of the biobank-related papers in the process. We start the overlap assessment with a naive analysis by directly considering all 6,899 publications from the NHGRI-EBI catalog, finding that 2,602 GWAS publications, or 38\% of the total, are also included in our publication data. In our dataset, these publications collectively mention 486 biobanks, the most mentioned by far being the UK Biobank with 748 publications, followed by FHS (169), Wellcome Trust Case Control Consortium (166), and the Rotterdam Study (156).

To measure our dataset's breadth in the reported mentions by individual biobank, we compared them to the reported numbers found in the NHGRI-EBI Catalog. Specifically, we extracted the number of publications that each biobank in the catalog is referenced as a data source and compared it to the number of mentions we find for that biobank across the catalog's publications.  We calculate these numbers for 89 biobanks having the same name in both datasets, finding that our dataset identifies, on average, 10.3 more publication-biobank associations than NHGRI-EBI. The UK Biobank is found in 748 publications in our dataset, but only in 485 publications of the reference catalog, a difference showing a much higher breadth in our dataset for this Biobank. The 263 publications mentioning the UK Biobank that do not list this biobank in the reference catalog have `NR' (230) missing values for the column `cohort' (33), suggesting that authors of the GWAS studies did not include the biobank source in the NHGRI-EBI Catalog. On average, the 72 biobanks better or equally covered in our dataset have 14.4 missing publications in the reference catalog. On the other hand, the 17 biobanks that are better covered in the NHGRI-EBI Catalog have 7.1 missing publications in our dataset.

Removing publications without a biobank assigned (cohort values 'NR', 'other', or 'NA'), our dataset covers 60\% of the 727 remaining publications in the GWAS catalog. These results show a good coverage of our dataset not without its limitations: while not all of the GWAS publications not covered (40\%) may have a population biobank as a data source, we miss mentions in the full-text of the paper. To have an approximate of what the percentage of missing mentions due to our lack of access to the full-text of a publication, we consider the 485 publications in the GWAS catalog identifying the UK biobank as their source. From these, 129 are not covered in our dataset, suggesting that 26\% of mentions of the UK Biobank may only be found in the body of the manuscript (excluding title, abstract, and acknowledgements). The data including the number of covered and missing publications is \url{data/database/meta/coverage.csv} and the code for the coverage analysis at \url{python/cohort_metadata/gwas_papers.py}.

\section{Co-citation network}\label{SI:network}

The biobank co-citation network of a biobank is built on top of the publications mentioning biobanks. The nodes of the network correspond to biobanks, and two biobanks are connected by an edge if their corresponding publications are often cited together. We operationalize the citations of a biobank as the number of articles citing the biobank's articles where it is mentioned. Mathematically, let $C_i = i_1, i_2, \dots i_n$ be the citing articles of biobank $i$, and $C_j = j_1, j_2, \dots j_m$ the citing articles of biobank $j$, then the union $C_i \cap C_j$ is composed by the articles citing both biobanks $i$ and $j$. For each biobank $i$, we only consider its most similar biobank by adding a single edge to the biobank $j$ such that $|C_i \cap C_j| = \max_x {|C_i \cap C_x|}$, for each biobank $x \neq i$. In other words, each biobank $i$ is connected by an edge to the biobank $j$ that is more often cited together with $i$.

\section{Hidden Citations}\label{SI:hidden-citations}

Hidden citations are a measure of the hidden impact of a biobank based on its reference papers~\cite{meng_hidden_2023}, and refer to unambiguous allusions made to a biobank without a corresponding citation. The idea is rooted in eponyms~\cite{cabanac_2014_eponym}, informal citations~\cite{marx2009informal}, and ``obliteration by incorperation''~\cite{mccain2012obliteration}. In the context of biobanks, these allusions appear as implicit citations based on contextual mentions of a biobank in the text of a scientific paper. A hidden citation to a biobank is defined as the difference between the explicit citations from papers mentioning the biobank and its total number of mentions.

\subsection{Reference papers}\label{SI:reference_papers}
In order to compute hidden citations, we first identify the foundational or reference papers of each biobank from the pool of papers that mention it. We identify a paper $x$ a reference paper of a biobank if $x$ mentions the biobank on its title or abstract and either (1) $x$ is the first publication mentioning the biobank; (2) a paper $y$ references $x$ with probability $P(x|y=m) > 0.2 + P(x)$, where $P(x|y=m)$ is the probability that $y$ references $x$ given that $y$ mentions the biobank and $P(x)$ is the probability that $x$ is cited by any paper; or (3) the title of $x$ contains the words `design', `baseline characteristics', or `rationale'. We identify at least one reference paper for 500 biobanks. The code to compute the missing citations is found at \url{python/iumpact/recognition/foundational.py}.

\subsection{Biobank Teams}\label{SI:biobank-teams}
The team of a biobank is composed of the union of all authors listed in one of its reference papers (Section~\ref{SI:reference_papers}). The data can be found at \url{data/expansion/database/cohort_impact/reference_papers.csv}.

\section{Null model}\label{SI:null-model}

In order to test the significance of the results showing that the impact of biobanks is local in terms of affiliation, country, and co-authorship, we create a null model in which we randomize the citing papers of each biobank such that the biobank remains with the same number of citations but with a random set of articles citing it. Once we have a randomized version of the citation network, where citing papers are randomized, we compute the mean share of in-house citations (coming from the same affiliation), the share of same-country citations, and the share of citations containing a collaborator of the biobank's team, and the share of citing papers containing its lead-author. We repeat this process 100 times and then use a z-test comparing the real value of each measure with its respective sequence of 100 randomized values to obtain a $p$-value.

\section{Disease Impact}\label{SI:disease-impact}
Based on MeSH terms, we link papers with diseases they have studied. For each disease, we count the total number of related articles per year for the period between 2000 and 2013. The number of articles in the sample is 10,486,605, studying 4,982 diseases. To have an idea of the disease impact of each biobank, we compute the ratio of publications relative to the total number of articles published for that disease in each year when the biobank was active. We limit the analysis to diseases with at least 10 publications per year in the global dataset. The resulting dataset then contains the proportion of articles related to each disease where the biobank is mentioned.

The last impact area of our proposed BIF is disease. This metric has three impact components of disease based on biobank publications: disease scope, disease depth, and rare disease (SI~\ref{SI:disease-impact}). The disease scope of the biobank is based on the number of different conditions studied from impacted papers relative to the total studies published for those diseases since the inception of the biobank, the UK Biobank leads this metric with 731 conditions studied, followed by NHANES (617), WHI, and the Telethon Network of Genetic Biobanks (TNGB, 199). The disease depth is based on the number of studies on a condition mentioning the biobank, relative to all publications on that condition, here the Avon Longitudinal Study of Parents and Children (ALSPAC) leads for environmental illness (14\% of all publications in 2020, followed by the Human Microbiome Project (HMP) for neoplastic processes (9\% of all publications in 2017), and the China National Genebank database (CNGBdb) for agricultural workers' diseases (7\% of publications in 2021). Finally, rare disease impact is a measure of the number of biobank publications on rare diseases, relative to all biobank publications, and is led by the St. Jude Lifetime Cohort Study (St. Jude LIFE) with 101 papers out of 104 on rare diseases, followed by the Swiss Childhood Cancer Survivor Study (SCCSS, 96\%), the NCI Childhood Cancer Survivor Study (CSS, 94\%), and the Ovarian Cancer Association Consortium database (OCAC, 88\%).

\subsection{MeSH diseases}\label{SI:mesh-diseases}

We use the definition of diseases from MeSH and operationalize a MeSH term as a disease if it is under the 'C' category of the MeSH tree. To compute the disease impact, we only consider articles published after the year 2000 mentioning a biobank, resulting in a total of 156,330 papers, we identify 75,269 as papers related to a condition (48\% of the total of papers mentioning 985 biobanks in the given time period). 

\subsection{UK Clinical Research Collaboration: UKCRC Diseases}\label{SI:UKCRC}
The HRCS system is used by a large number of health research funders in the UK and is subdivided into the Research Activity Classifications (RAC) and Health Categories (HC).

\subsection{Research, Condition, and Disease Categorization (RCDC)}\label{SI:RCDC}
This categorization system is used by the NIH to report to Congress and is a biomedical system consisting of 237 categories, some of which are very specific in the topic (e.g. ``ataxia-telangiectasia"), and others more general (e.g. ``neuroscience").

\subsection{Broad disease impact}
For each biobank, we define its broad disease impact as the sum across diseases studied with the biobank data of the proportion of papers mentioning one of the 985 biobanks each year.

\subsection{Specific disease impact}
To account for the impact of biobanks focusing on a single disease, we define the specific disease impact as the highest proportion of articles published mentioning the biobank for a given disease relative to the total number of articles published for that disease during the active years of the biobank. Here, we limit our study to diseases in the second classification level of MeSH with at least 100 publications to account only for diseases with an established interest from the scientific community, resulting in 180 specific diseases related to 632 biobanks.

\section{Mentions data}\label{SI:mentions-data}
For each document mentioning a biobank, we collect the relevant metadata to implement each of the impact measures developed.

\subsection{Grants data}\label{SI:grant-data}
For each grant, we extract its NIH activity code, the funder, funder amount, and disease categories (based on MeSH, RCDC, and HRCS). The NIH activity codes are then used to identify training, collaboration, and prestigious grants.

\subsection{Patents data}\label{SI:patent-data}
For each patent, we identify its Cooperative Patent Classification (CPC) and its disease categories.

\subsection{Clinical trials data}\label{SI:clinical-trials-data}
From each clinical trial, we extract its disease categories, as well as its assignee, and its institutional type.

\subsection{Public policies data}\label{SI:public-policy-data}
For each public policy document, we extract its country of publication based on the affiliation of the institution implementing the policy, and the type of its implementing institution.

\section{Biobank data}\label{SI:biobank-data}

Biobanks differ in the scope and kind of data they offer. A Biobank may focus on a particular physical sample, e.g., brain tissue, while another may offer multiple sample types and questionnaire information from its participants. Genetic biomarkers, those gene variants related to higher risk or diagnosis of a disease, are among the most coveted goals of cohort studies and require DNA data from biobanks. The kind of genetic data, however, can define the methodology used by the study, however, and is therefore highly relevant. Similarly, medical records can be used to trace risk factors to multiple diseases and conditions suffered by the biobank's participants, complementing its data. Here, we assess the type of data available from the biobank across several factors including, genetic data type, environmental data, follow-up data, medical health records, and disease-specific registry data. In order to identify the type of data available from the biobank, we looked at the MeSH classification terms of the articles mentioning the biobank. If multiple biobank papers are Genome-Wide Association Studies (GWAS), it is highly likely the biobank contains GWAS data available. Here we present the different MeSH terms we used to associate each data type. The complete set of MeSH terms and tree numbers used can be found at data/mesh\_and\_meta/mesh\_terms.md.

\subsection{Genetic data}
In order to assess whether a biobank has genetic data available, we looked for the following MeSH terms in the articles mentioning it: Genome-wide Association Study, Whole Genome Sequencing, Genetic Association Studies, Genetic Techniques, as well as any MeSH term under the tree classification of Genetic Predisposition to Disease (MeSH tree number C23.550.291.906). We then sub-classified the DNA data on GWAS data and Whole Genome Sequencing. The code can be found at python/expansion/data/dna.py.

\subsection{Environment data}
We used the sub-branches of each of the following MeSH terms to identify papers with environmental data: Gene-environment Interaction, Environmental Exposure, Environmental Biomarkers, Environmental Indicators, Environment, and Environment Design. The code can be found at python/expansion/data/environment.py.

\subsection{Follow-up data}
Often, a biobank's cohort is followed through time, and samples and data are collected on each follow-up. To identify follow-up studies of a biobank, we used the MeSH term `Follow-Up Studies' (D005500). The code can be found at python/expansion /data/follow\_up.py.

\subsection{Medical records}
We used the MeSH terms under the tree classification of `Medical Records' (MeSH tree ID E05.318.308.940.968). The code can be found at python/expansion /data/medical\_records.py.

\subsection{Registries and disease-specific data}
In order to look for registries and other types of disease-specific data such as the one coming from panels, we used the following MeSH terms: Registries (E05.318.308.970), National Program of Cancer Registries (I01.409.418.750.600.650.200.760), and Mass Screening (E01.370.500). The code can be found at python/expansion /data/registries \_disease\_specific.py.

\subsection{Surveys and Questionnaires}
Some studies rely on data from surveys and questionnaires to build a social profile of the participants of the biobank. To identify those biobanks with survey and questionnaire data we identify the papers under the mesh term Surveys and Questionnaires (E05.318.308.980). The code can be found at surveys\_and\_questionnaires.py.

\subsection{Age groups}
Most biobank cohorts are designed with an age group in mind, usually defined by the first wave of data collected from its participants. In order to identify the age group of a biobank, we identified all the age groups under the MeSH classification of Age Groups (M01.060) and classified it into 7 categories depending on the tree number of the age group, namely: Middle Aged (M01.060.116.630), aged (M01.060.116.100), young adult (M01.060.116.815), adolescent (M01.060.057), infant (M01.060.703), child (M01.060.406), and birth (M01.060.261). The code can be found at persons.py.

\subsection{Country of origin}
We collected the country of origin of the cohort from each biobank by identifying the MeSH terms related to countries in their publications. This includes every term under the MeSH tree classification of Geographic Locations (Z01), except for cities (Z01.433). We also consider the countries where the authors of the papers are affiliated based on their institution on each paper. The code can be found at countries.py.

\subsection{Open data index}\label{si:open-data-index}

In order to assess the level of \emph{data openness} of a biobank, we consider three factors related to the papers mentioning the biobank: the percentage of papers led by the top corresponding author, the percentage of papers co-authored by the top 10 authors, and the percentage of papers with at least one author affiliated to the top affiliation. We then define the `open data index' of a biobank as 100 minus the mean of these percentages.

\subsection{Cohort size}

The cohort size of a biobank, i.e., the number of participants or individuals included for sampling, surveys, and/or follow-up, may be of interest to researchers looking to access its data. Although Biobanks usually publish the size of their cohort on their design or foundational papers, it may be subject to change in subsequent updates of the biobank, or it may be a target number. Oftentimes, moreover, researchers may be interested in a sub-sample of the cohort depending on multiple factors such as the disease of interest, the analytical method of their study, or the type of data they need. In order to have an idea of the cohort size that is used across the research papers that make use of a biobank's data, we look for the self-disclosed sample sizes appearing in the abstracts of such studies. This approach makes use of papers mentioning a biobank, resulting in a distribution of sample sizes used across time and topics. In order to identify sample sizes we use a regular expression matching numbers followed by a keyword (code available at SQL/expansion/cohort\_size.sql), and to filter out those papers that may be using more than one biobank, we only consider articles mentioning one biobank in our sample. For each biobank, we then consider the 90th percentile as its cohort size to consider the larger sub-samples of the biobank and to avoid outliers that may represent noisy values. The code to create the table with the sample size of each biobank is found at \url{python/data/cohort_size.py}.

\subsection{Population vs.\ health based biobanks}

Biobank data is retrieved from a cohort of participants that may be recruited from a population defined by a geographical location, a line of work, or any other common factor, or it can be based on patients (dead or alive) coming from a hospital or clinic and usually sharing a disease~\cite{baker_building_2012}. The border between population and patients, however, is not always clear. This is the case when biobanks recruit individuals from a general population that are patients are local clinics, like in the UK Biobank. Indeed, often participants are invited to participate in the Biobank's cohort from a clinic or a hospital, not because they are patients there, but because it is strategic or convenient (e.g., to link their medical records to their samples).

To assess whether a biobank's design had a population or a group of patients in mind we identified several keywords from the abstracts of the papers mentioning the biobank, including `population-based', `patients', `hospital', `health-based', and `clinic'. To classify the cohort type of the biobank, we then computed the ratio of papers with each keyword and assigned the type with the highest proportion of papers. Under this approach, the UK Biobank, even if around 10\% of its papers contain the word `patients', is classified as a population-based biobank as 29\% of its articles contain the keyword `population'. The code can be found at python/expansion/database/cohort\_type.py.

\section{Biobank Impact Factor}\label{SI:BIF}
We considered two factors to calculate the Biobank Impact Factor (BIF): mentions and disease impact. To calculate the mention impact of a biobank, we standardize its number of mentions for each document type separately (i.e., publications, grants, patents, clinical trials, and public policy documents). For each document type, standardization is done by taking the number of mentions $m$ of a biobank and returning $(m - \mu) / \sigma$, where $\mu$ and $\sigma$ is the mean and standard deviation over all biobanks. Once every value is standardized, we sum the values of each biobank across all document types, this is the mention impact. To consider the advantage in mentions of older biobanks, where the number of mentions may be related to the time period they have been available, we normalize mention impact by the number of years the biobank have been mentioned in scientific publications.

This mention impact takes into account the overall presence of biobanks across science, innovation, and policy, as well as their lack thereof. Because mentions are standardized by document type, a biobank can have a negative value that results in a `penalty' for its impact. The BIF scores higher balanced biobanks, those having an impact, even if modest, across most sectors. The code to compute mention impact is found in \url{python/impact/cohort_impact_factor/impact_factor.py}.

Another factor that we consider for BIF is the scope and depth of research produced by each biobank (SI~\ref{SI:disease-impact}). This metric considers the number of publications each biobank helps produce for a disease, compared to the total number of publications about the disease (depth); as well as the total number of diseases studied on publications mentioning the biobank (scope). Additionally, disease impact considers the impact the biobank has on rare diseases, measured by the number of mentioning publications on a rare disease, relative to the all other biobanks. This ensures that biobanks contributing to a single disease, and those contributing to rare diseases, are considered in the BIF. The code to compute disease impact is found in \url{python/impact/cohort_impact_factor/disease_impact.py}.

The code to compute the BIF is found in \url{python/impact/cohort_impact_factor/target.py}.

\section{Generalized Linear Model}\label{glm}

We used a Generalized Linear Model (GLM) to measure the relation between biobank data characteristics and BIF. The dependent variable is then BIF and the 14 independent variables were all binary values depending on the data characteristics of each biobank (Table~\ref{SItable:model}). Specifically, we used a Gaussian family GLM with identity function linked to the data, using the statsmodels library in Python. The model was applied to a set of 468 biobanks for which we could identify each feature (SI~\ref{SI:biobank-data}). For each dependent feature we obtained its coefficient and associated $p$-value, labeling it as significant if $p$-value $0.05 / n$, where $n$ is the number of features in the model (Bonferroni correction for each independent feature). The model presents a an $R$-square metric of 0.4 and log-likelihood of 862.65. The code to fit the data to the GLM and to obtain the significant variables is at \url{python/stat_models/glm.py}.

\begin{center}
\begin{tabular}{lclc}
\toprule
\textbf{Dep. Variable:}   &      target      & \textbf{  No. Observations:  } &      468    \\
\textbf{Model:}           &       GLM        & \textbf{  Df Residuals:      } &      453    \\
\textbf{Model Family:}    &     Gaussian     & \textbf{  Df Model:          } &       14    \\
\textbf{Link Function:}   &     Identity     & \textbf{  Scale:             } & 0.0015157   \\
\textbf{Method:}          &       IRLS       & \textbf{  Log-Likelihood:    } &    862.65   \\
\textbf{Date:}            & Mon, 15 Apr 2024 & \textbf{  Deviance:          } &   0.68663   \\
\textbf{Time:}            &     10:02:51     & \textbf{  Pearson chi2:      } &    0.687    \\
\textbf{No. Iterations:}  &        3         & \textbf{  Pseudo R-squ. (CS):} &   0.4097    \\
\textbf{Covariance Type:} &    nonrobust     & \textbf{                     } &             \\
\bottomrule
\label{SItable:model}
\end{tabular}

\begin{tabular}{lcccccc}
                          & \textbf{coef} & \textbf{std err} & \textbf{z} & \textbf{P$> |$z$|$} & \textbf{[0.025} & \textbf{0.975]}  \\
\midrule
\textbf{const}            &      -0.0186  &        0.009     &    -2.102  &         0.036        &       -0.036    &       -0.001     \\
\textbf{dna}              &       0.0002  &        0.005     &     0.039  &         0.969        &       -0.010    &        0.011     \\
\textbf{gwas}             &       0.0024  &        0.005     &     0.489  &         0.625        &       -0.007    &        0.012     \\
\textbf{whole\_genome}    &       0.0286  &        0.006     &     4.752  &         0.000        &        0.017    &        0.040     \\
\textbf{gene\_env}        &       0.0267  &        0.005     &     5.346  &         0.000        &        0.017    &        0.037     \\
\textbf{follow\_up}       &       0.0144  &        0.005     &     2.900  &         0.004        &        0.005    &        0.024     \\
\textbf{medical\_records} &       0.0187  &        0.004     &     4.210  &         0.000        &        0.010    &        0.027     \\
\textbf{surveys}          &      -0.0049  &        0.008     &    -0.608  &         0.543        &       -0.021    &        0.011     \\
\textbf{registries}       &       0.0077  &        0.004     &     1.857  &         0.063        &       -0.000    &        0.016     \\
\textbf{population}       &      -0.0010  &        0.005     &    -0.221  &         0.825        &       -0.010    &        0.008     \\
\textbf{large\_cohort}    &       0.0048  &        0.006     &     0.796  &         0.426        &       -0.007    &        0.017     \\
\textbf{open\_data\_high} &       0.0345  &        0.007     &     5.196  &         0.000        &        0.021    &        0.048     \\
\textbf{open\_data\_low}  &      -0.0051  &        0.008     &    -0.618  &         0.537        &       -0.021    &        0.011     \\
\textbf{PI\_high\_impact} &       0.0198  &        0.006     &     3.079  &         0.002        &        0.007    &        0.032     \\
\textbf{PI\_low\_impact}  &      -0.0092  &        0.007     &    -1.336  &         0.182        &       -0.023    &        0.004     \\
\bottomrule
\end{tabular}
\end{center}

\printbibliography

@article{wijmenga_importance_2018,
	title = {The importance of cohort studies in the post-{GWAS} era},
	volume = {50},
	copyright = {2018 The Author(s)},
	issn = {1546-1718},
	url = {https://www.nature.com/articles/s41588-018-0066-3},
	doi = {10.1038/s41588-018-0066-3},
	language = {en},
	number = {3},
	urldate = {2024-06-11},
	journal = {Nature Genetics},
	author = {Wijmenga, Cisca and Zhernakova, Alexandra},
	month = mar,
	year = {2018},
	note = {Publisher: Nature Publishing Group},
	pages = {322--328},
}

@article{sollis_nhgri-ebi_2023,
	title = {The {NHGRI}-{EBI} {GWAS} {Catalog}: knowledgebase and deposition resource},
	volume = {51},
	issn = {0305-1048},
	shorttitle = {The {NHGRI}-{EBI} {GWAS} {Catalog}},
	url = {https://doi.org/10.1093/nar/gkac1010},
	doi = {10.1093/nar/gkac1010},
	number = {D1},
	urldate = {2024-06-11},
	journal = {Nucleic Acids Research},
	author = {Sollis, Elliot and Mosaku, Abayomi and Abid, Ala and Buniello, Annalisa and Cerezo, Maria and Gil, Laurent and Groza, Tudor and Güneş, Osman and Hall, Peggy and Hayhurst, James and Ibrahim, Arwa and Ji, Yue and John, Sajo and Lewis, Elizabeth and MacArthur, Jacqueline A L and McMahon, Aoife and Osumi-Sutherland, David and Panoutsopoulou, Kalliope and Pendlington, Zoë and Ramachandran, Santhi and Stefancsik, Ray and Stewart, Jonathan and Whetzel, Patricia and Wilson, Robert and Hindorff, Lucia and Cunningham, Fiona and Lambert, Samuel A and Inouye, Michael and Parkinson, Helen and Harris, Laura W},
	month = jan,
	year = {2023},
	pages = {D977--D985},
}

@article{lerche_methods_2015,
	title = {Methods in {Neuroepidemiology} {Characterization} of {European} {Longitudinal} {Cohort} {Studies} in {Parkinson}'s {Disease} - {Report} of the {JPND} {Working} {Group} {BioLoC}-{PD}},
	volume = {45},
	issn = {0251-5350},
	url = {https://doi.org/10.1159/000439221},
	doi = {10.1159/000439221},
	number = {4},
	urldate = {2024-06-04},
	journal = {Neuroepidemiology},
	author = {Lerche, Stefanie and Liepelt-Scarfone, Inga and Alves, Guido and Barone, Paolo and Behnke, Stefanie and Ben-Shlomo, Yoav and Berendse, Henk and Burn, David and Dodel, Richard and Grosset, Donald and Heinzel, Sebastian and Hu, Michele and Kasten, Meike and Krüger, Rejko and Maetzler, Walter and Moccia, Marcello and Mollenhauer, Brit and Oertel, Wolfgang and Roeben, Benjamin and Sünkel, Ulrike and Walter, Uwe and Wirdefeldt, Karin and Berg, Daniela},
	month = nov,
	year = {2015},
	pages = {282--297},
}

@article{tryka_ncbis_2014,
	title = {{NCBI}’s {Database} of {Genotypes} and {Phenotypes}: {dbGaP}},
	volume = {42},
	issn = {0305-1048},
	shorttitle = {{NCBI}’s {Database} of {Genotypes} and {Phenotypes}},
	url = {https://doi.org/10.1093/nar/gkt1211},
	doi = {10.1093/nar/gkt1211},
	number = {D1},
	urldate = {2024-06-04},
	journal = {Nucleic Acids Research},
	author = {Tryka, Kimberly A. and Hao, Luning and Sturcke, Anne and Jin, Yumi and Wang, Zhen Y. and Ziyabari, Lora and Lee, Moira and Popova, Natalia and Sharopova, Nataliya and Kimura, Masato and Feolo, Michael},
	month = jan,
	year = {2014},
	pages = {D975--D979},
}

@article{bergeron_fostering_2018,
	title = {Fostering population-based cohort data discovery: {The} {Maelstrom} {Research} cataloguing toolkit},
	volume = {13},
	issn = {1932-6203},
	shorttitle = {Fostering population-based cohort data discovery},
	url = {https://journals.plos.org/plosone/article?id=10.1371/journal.pone.0200926},
	doi = {10.1371/journal.pone.0200926},
	language = {en},
	number = {7},
	urldate = {2024-06-04},
	journal = {PLOS ONE},
	author = {Bergeron, Julie and Doiron, Dany and Marcon, Yannick and Ferretti, Vincent and Fortier, Isabel},
	month = jul,
	year = {2018},
	note = {Publisher: Public Library of Science},
	pages = {e0200926},
}

@article{smith-warner_methods_2006,
	title = {Methods for {Pooling} {Results} of {Epidemiologic} {Studies}: {The} {Pooling} {Project} of {Prospective} {Studies} of {Diet} and {Cancer}},
	volume = {163},
	issn = {0002-9262},
	shorttitle = {Methods for {Pooling} {Results} of {Epidemiologic} {Studies}},
	url = {https://doi.org/10.1093/aje/kwj127},
	doi = {10.1093/aje/kwj127},
	number = {11},
	urldate = {2024-06-04},
	journal = {American Journal of Epidemiology},
	author = {Smith-Warner, Stephanie A. and Spiegelman, Donna and Ritz, John and Albanes, Demetrius and Beeson, W. Lawrence and Bernstein, Leslie and Berrino, Franco and van den Brandt, Piet A. and Buring, Julie E. and Cho, Eunyoung and Colditz, Graham A. and Folsom, Aaron R. and Freudenheim, Jo L. and Giovannucci, Edward and Goldbohm, R. Alexandra and Graham, Saxon and Harnack, Lisa and Horn-Ross, Pamela L. and Krogh, Vittorio and Leitzmann, Michael F. and McCullough, Marjorie L. and Miller, Anthony B. and Rodriguez, Carmen and Rohan, Thomas E. and Schatzkin, Arthur and Shore, Roy and Virtanen, Mikko and Willett, Walter C. and Wolk, Alicja and Zeleniuch-Jacquotte, Anne and Zhang, Shumin M. and Hunter, David J.},
	month = jun,
	year = {2006},
	pages = {1053--1064},
}

@article{kennedy_cancer_2016,
	title = {The {Cancer} {Epidemiology} {Descriptive} {Cohort} {Database}: {A} {Tool} to {Support} {Population}-{Based} {Interdisciplinary} {Research}},
	volume = {25},
	issn = {1055-9965},
	shorttitle = {The {Cancer} {Epidemiology} {Descriptive} {Cohort} {Database}},
	url = {https://doi.org/10.1158/1055-9965.EPI-16-0412},
	doi = {10.1158/1055-9965.EPI-16-0412},
	number = {10},
	urldate = {2024-06-04},
	journal = {Cancer Epidemiology, Biomarkers \& Prevention},
	author = {Kennedy, Amy E. and Khoury, Muin J. and Ioannidis, John P.A. and Brotzman, Michelle and Miller, Amy and Lane, Crystal and Lai, Gabriel Y. and Rogers, Scott D. and Harvey, Chinonye and Elena, Joanne W. and Seminara, Daniela},
	month = oct,
	year = {2016},
	pages = {1392--1401},
}

@article{vrijheid_european_2012,
	title = {European {Birth} {Cohorts} for {Environmental} {Health} {Research}},
	volume = {120},
	url = {https://ehp.niehs.nih.gov/doi/full/10.1289/ehp.1103823},
	doi = {10.1289/ehp.1103823},
	number = {1},
	urldate = {2024-06-04},
	journal = {Environmental Health Perspectives},
	author = {Vrijheid, Martine and Casas, Maribel and Bergström, Anna and Carmichael, Amanda and Cordier, Sylvaine and Eggesbø, Merete and Eller, Esben and Fantini, Maria P. and Fernández, Mariana F. and Fernández-Somoano, Ana and Gehring, Ulrike and Grazuleviciene, Regina and Hohmann, Cynthia and Karvonen, Anne M. and Keil, Thomas and Kogevinas, Manolis and Koppen, Gudrun and Krämer, Ursula and Kuehni, Claudia E. and Magnus, Per and Majewska, Renata and Andersen, Anne-Marie Nybo and Patelarou, Evridiki and Petersen, Maria Skaalum and Pierik, Frank H. and Polanska, Kinga and Porta, Daniela and Richiardi, Lorenzo and Santos, Ana Cristina and Slama, Rémy and Sram, Radim J. and Thijs, Carel and Tischer, Christina and Toft, Gunnar and Trnovec, Tomáš and Vandentorren, Stephanie and Vrijkotte, Tanja G.M. and Wilhelm, Michael and Wright, John and Nieuwenhuijsen, Mark},
	month = jan,
	year = {2012},
	note = {Publisher: Environmental Health Perspectives},
	pages = {29--37},
}

@article{bauermeister_dementias_2020,
	title = {The {Dementias} {Platform} {UK} ({DPUK}) {Data} {Portal}},
	volume = {35},
	issn = {1573-7284},
	url = {https://doi.org/10.1007/s10654-020-00633-4},
	doi = {10.1007/s10654-020-00633-4},
	language = {en},
	number = {6},
	urldate = {2024-06-04},
	journal = {European Journal of Epidemiology},
	author = {Bauermeister, Sarah and Orton, Christopher and Thompson, Simon and Barker, Roger A. and Bauermeister, Joshua R. and Ben-Shlomo, Yoav and Brayne, Carol and Burn, David and Campbell, Archie and Calvin, Catherine and Chandran, Siddharthan and Chaturvedi, Nishi and Chêne, Geneviève and Chessell, Iain P. and Corbett, Anne and Davis, Daniel H. J. and Denis, Mike and Dufouil, Carole and Elliott, Paul and Fox, Nick and Hill, Derek and Hofer, Scott M. and Hu, Michele T. and Jindra, Christoph and Kee, Frank and Kim, Chi-Hun and Kim, Changsoo and Kivimaki, Mika and Koychev, Ivan and Lawson, Rachael A. and Linden, Gerry J. and Lyons, Ronan A. and Mackay, Clare and Matthews, Paul M. and McGuiness, Bernadette and Middleton, Lefkos and Moody, Catherine and Moore, Katrina and Na, Duk L. and O’Brien, John T. and Ourselin, Sebastien and Paranjothy, Shantini and Park, Ki-Soo and Porteous, David J. and Richards, Marcus and Ritchie, Craig W. and Rohrer, Jonathan D. and Rossor, Martin N. and Rowe, James B. and Scahill, Rachael and Schnier, Christian and Schott, Jonathan M. and Seo, Sang W. and South, Matthew and Steptoe, Matthew and Tabrizi, Sarah J. and Tales, Andrea and Tillin, Therese and Timpson, Nicholas J. and Toga, Arthur W. and Visser, Pieter-Jelle and Wade-Martins, Richard and Wilkinson, Tim and Williams, Julie and Wong, Andrew and Gallacher, John E. J.},
	month = jun,
	year = {2020},
	pages = {601--611},
}

@article{wang_improved_2012,
	title = {Improved imputation of common and uncommon {SNPs} with a new reference set},
	volume = {44},
	copyright = {2012 Springer Nature America, Inc.},
	issn = {1546-1718},
	url = {https://www.nature.com/articles/ng.1044},
	doi = {10.1038/ng.1044},
	language = {en},
	number = {1},
	urldate = {2024-06-04},
	journal = {Nature Genetics},
	author = {Wang, Zhaoming and Jacobs, Kevin B. and Yeager, Meredith and Hutchinson, Amy and Sampson, Joshua and Chatterjee, Nilanjan and Albanes, Demetrius and Berndt, Sonja I. and Chung, Charles C. and Diver, W. Ryan and Gapstur, Susan M. and Teras, Lauren R. and Haiman, Christopher A. and Henderson, Brian E. and Stram, Daniel and Deng, Xiang and Hsing, Ann W. and Virtamo, Jarmo and Eberle, Michael A. and Stone, Jennifer L. and Purdue, Mark P. and Taylor, Phil and Tucker, Margaret and Chanock, Stephen J.},
	month = jan,
	year = {2012},
	note = {Publisher: Nature Publishing Group},
	pages = {6--7},
}

@article{ouellette_p3g_2014,
	title = {{P3G} — 10 years of toolbuilding: {From} the population biobank to the clinic},
	volume = {3},
	issn = {2212-0661},
	shorttitle = {{P3G} — 10 years of toolbuilding},
	url = {https://www.ncbi.nlm.nih.gov/pmc/articles/PMC4882047/},
	doi = {10.1016/j.atg.2014.04.004},
	number = {2},
	urldate = {2024-06-04},
	journal = {Applied \& Translational Genomics},
	author = {Ouellette, Sylvie and Tassé, Anne Marie},
	month = apr,
	year = {2014},
	pmid = {27275412},
	pmcid = {PMC4882047},
	pages = {36--40},
}

@article{van_der_velde_molgenis_2019,
	title = {{MOLGENIS} research: advanced bioinformatics data software for non-bioinformaticians},
	volume = {35},
	issn = {1367-4803},
	shorttitle = {{MOLGENIS} research},
	url = {https://doi.org/10.1093/bioinformatics/bty742},
	doi = {10.1093/bioinformatics/bty742},
	number = {6},
	urldate = {2024-06-04},
	journal = {Bioinformatics},
	author = {van der Velde, K Joeri and Imhann, Floris and Charbon, Bart and Pang, Chao and van Enckevort, David and Slofstra, Mariska and Barbieri, Ruggero and Alberts, Rudi and Hendriksen, Dennis and Kelpin, Fleur and de Haan, Mark and de Boer, Tommy and Haakma, Sido and Stroomberg, Connor and Scholtens, Salome and van de Geijn, Gert-Jan and Festen, Eleonora A M and Weersma, Rinse K and Swertz, Morris A},
	month = mar,
	year = {2019},
	pages = {1076--1078},
}

@article{liggins_international_2014,
	title = {International {Alzheimer}'s {Disease} {Research} {Portfolio} ({IADRP}) aims to capture global {Alzheimer}'s disease research funding},
	volume = {10},
	copyright = {© 2014 The Alzheimer's Association},
	issn = {1552-5279},
	url = {https://onlinelibrary.wiley.com/doi/abs/10.1016/j.jalz.2013.12.013},
	doi = {10.1016/j.jalz.2013.12.013},
	language = {en},
	number = {3},
	urldate = {2024-06-04},
	journal = {Alzheimer's \& Dementia},
	author = {Liggins, Charlene and Snyder, Heather M. and Silverberg, Nina and Petanceska, Suzana and Refolo, Lorenzo M. and Ryan, Laurie and Carrillo, Maria C.},
	year = {2014},
	note = {\_eprint: https://onlinelibrary.wiley.com/doi/pdf/10.1016/j.jalz.2013.12.013},
	pages = {405--408},
}

@article{bose_data_2022,
	title = {Data and sample sharing as an enabler for large-scale biomarker research and development: {The} {EPND} perspective},
	volume = {13},
	issn = {1664-2295},
	shorttitle = {Data and sample sharing as an enabler for large-scale biomarker research and development},
	url = {https://www.frontiersin.org/journals/neurology/articles/10.3389/fneur.2022.1031091/full},
	doi = {10.3389/fneur.2022.1031091},
	language = {English},
	urldate = {2024-06-04},
	journal = {Frontiers in Neurology},
	author = {Bose, Niranjan and Brookes, Anthony J. and Scordis, Phil and Visser, Pieter Jelle},
	month = nov,
	year = {2022},
	note = {Publisher: Frontiers},
}

@article{giffen_providing_2017,
	title = {Providing researchers with online access to {NHLBI} biospecimen collections: {The} results of the first six years of the {NHLBI} {BioLINCC} program},
	volume = {12},
	issn = {1932-6203},
	shorttitle = {Providing researchers with online access to {NHLBI} biospecimen collections},
	url = {https://journals.plos.org/plosone/article?id=10.1371/journal.pone.0178141},
	doi = {10.1371/journal.pone.0178141},
	language = {en},
	number = {6},
	urldate = {2024-06-04},
	journal = {PLOS ONE},
	author = {Giffen, Carol A. and Wagner, Elizabeth L. and Adams, John T. and Hitchcock, Denise M. and Welniak, Lisbeth A. and Brennan, Sean P. and Carroll, Leslie E.},
	month = jun,
	year = {2017},
	note = {Publisher: Public Library of Science},
	pages = {e0178141},
}

@article{otlowski_biobanks_2020,
	title = {Biobanks {Information} {Paper} 2010},
	volume = {20},
	url = {https://search.informit.org/doi/abs/10.3316/ielapa.618168436076064},
	doi = {10.3316/ielapa.618168436076064},
	number = {1},
	urldate = {2024-04-19},
	journal = {Journal of Law, Information and Science},
	author = {Otlowski, Margaret and Nicol, Dianne and Stranger, Mark},
	month = dec,
	year = {2020},
	note = {Publisher: University of Tasmania - Faculty of Law},
	pages = {97--227},
}

@article{brumpton_hunt_2022,
	title = {The {HUNT} study: {A} population-based cohort for genetic research},
	volume = {2},
	issn = {2666979X},
	shorttitle = {The {HUNT} study},
	url = {https://linkinghub.elsevier.com/retrieve/pii/S2666979X22001422},
	doi = {10.1016/j.xgen.2022.100193},
	language = {en},
	number = {10},
	urldate = {2024-04-18},
	journal = {Cell Genomics},
	author = {Brumpton, Ben M. and Graham, Sarah and Surakka, Ida and Skogholt, Anne Heidi and Løset, Mari and Fritsche, Lars G. and Wolford, Brooke and Zhou, Wei and Nielsen, Jonas Bille and Holmen, Oddgeir L. and Gabrielsen, Maiken E. and Thomas, Laurent and Bhatta, Laxmi and Rasheed, Humaira and Zhang, He and Kang, Hyun Min and Hornsby, Whitney and Moksnes, Marta Riise and Coward, Eivind and Melbye, Mads and Giskeødegård, Guro F. and Fenstad, Jørn and Krokstad, Steinar and Næss, Marit and Langhammer, Arnulf and Boehnke, Michael and Abecasis, Gonçalo R. and Åsvold, Bjørn Olav and Hveem, Kristian and Willer, Cristen J.},
	month = oct,
	year = {2022},
	pages = {100193},
}

@article{petersen_imaging_2013,
	title = {Imaging in population science: cardiovascular magnetic resonance in 100,000 participants of {UK} {Biobank} - rationale, challenges and approaches},
	volume = {15},
	issn = {1532-429X},
	shorttitle = {Imaging in population science},
	url = {https://doi.org/10.1186/1532-429X-15-46},
	doi = {10.1186/1532-429X-15-46},
	language = {en},
	number = {1},
	urldate = {2024-04-15},
	journal = {Journal of Cardiovascular Magnetic Resonance},
	author = {Petersen, Steffen E. and Matthews, Paul M. and Bamberg, Fabian and Bluemke, David A. and Francis, Jane M. and Friedrich, Matthias G. and Leeson, Paul and Nagel, Eike and Plein, Sven and Rademakers, Frank E. and Young, Alistair A. and Garratt, Steve and Peakman, Tim and Sellors, Jonathan and Collins, Rory and Neubauer, Stefan},
	month = may,
	year = {2013},
	pages = {46},
}

@article{littlejohns_uk_2020,
	title = {The {UK} {Biobank} imaging enhancement of 100,000 participants: rationale, data collection, management and future directions},
	volume = {11},
	copyright = {2020 The Author(s)},
	issn = {2041-1723},
	shorttitle = {The {UK} {Biobank} imaging enhancement of 100,000 participants},
	url = {https://www.nature.com/articles/s41467-020-15948-9},
	doi = {10.1038/s41467-020-15948-9},
	language = {en},
	number = {1},
	urldate = {2024-04-14},
	journal = {Nature Communications},
	author = {Littlejohns, Thomas J. and Holliday, Jo and Gibson, Lorna M. and Garratt, Steve and Oesingmann, Niels and Alfaro-Almagro, Fidel and Bell, Jimmy D. and Boultwood, Chris and Collins, Rory and Conroy, Megan C. and Crabtree, Nicola and Doherty, Nicola and Frangi, Alejandro F. and Harvey, Nicholas C. and Leeson, Paul and Miller, Karla L. and Neubauer, Stefan and Petersen, Steffen E. and Sellors, Jonathan and Sheard, Simon and Smith, Stephen M. and Sudlow, Cathie L. M. and Matthews, Paul M. and Allen, Naomi E.},
	month = may,
	year = {2020},
	note = {Publisher: Nature Publishing Group},
	pages = {2624},
}

@article{fry_comparison_2017,
	title = {Comparison of {Sociodemographic} and {Health}-{Related} {Characteristics} of {UK} {Biobank} {Participants} {With} {Those} of the {General} {Population}},
	volume = {186},
	issn = {0002-9262},
	url = {https://doi.org/10.1093/aje/kwx246},
	doi = {10.1093/aje/kwx246},
	number = {9},
	urldate = {2024-04-14},
	journal = {American Journal of Epidemiology},
	author = {Fry, Anna and Littlejohns, Thomas J and Sudlow, Cathie and Doherty, Nicola and Adamska, Ligia and Sprosen, Tim and Collins, Rory and Allen, Naomi E},
	month = nov,
	year = {2017},
	pages = {1026--1034},
}

@article{bycroft_uk_2018,
	title = {The {UK} {Biobank} resource with deep phenotyping and genomic data},
	volume = {562},
	copyright = {2018 Springer Nature Limited},
	issn = {1476-4687},
	url = {https://www.nature.com/articles/s41586-018-0579-z.},
	doi = {10.1038/s41586-018-0579-z},
	language = {en},
	number = {7726},
	urldate = {2024-04-14},
	journal = {Nature},
	author = {Bycroft, Clare and Freeman, Colin and Petkova, Desislava and Band, Gavin and Elliott, Lloyd T. and Sharp, Kevin and Motyer, Allan and Vukcevic, Damjan and Delaneau, Olivier and O’Connell, Jared and Cortes, Adrian and Welsh, Samantha and Young, Alan and Effingham, Mark and McVean, Gil and Leslie, Stephen and Allen, Naomi and Donnelly, Peter and Marchini, Jonathan},
	month = oct,
	year = {2018},
	note = {Publisher: Nature Publishing Group},
	pages = {203--209},
}

@article{sudlow_uk_2015,
	title = {{UK} {Biobank}: {An} {Open} {Access} {Resource} for {Identifying} the {Causes} of a {Wide} {Range} of {Complex} {Diseases} of {Middle} and {Old} {Age}},
	volume = {12},
	issn = {1549-1676},
	shorttitle = {{UK} {Biobank}},
	url = {https://journals.plos.org/plosmedicine/article?id=10.1371/journal.pmed.1001779},
	doi = {10.1371/journal.pmed.1001779},
	language = {en},
	number = {3},
	urldate = {2024-04-14},
	journal = {PLOS Medicine},
	author = {Sudlow, Cathie and Gallacher, John and Allen, Naomi and Beral, Valerie and Burton, Paul and Danesh, John and Downey, Paul and Elliott, Paul and Green, Jane and Landray, Martin and Liu, Bette and Matthews, Paul and Ong, Giok and Pell, Jill and Silman, Alan and Young, Alan and Sprosen, Tim and Peakman, Tim and Collins, Rory},
	month = mar,
	year = {2015},
	note = {Publisher: Public Library of Science},
	pages = {e1001779},
}

@article{nagai_overview_2017,
	title = {Overview of the {BioBank} {Japan} {Project}: {Study} design and profile},
	volume = {27},
	shorttitle = {Overview of the {BioBank} {Japan} {Project}},
	doi = {10.1016/j.je.2016.12.005},
	number = {Supplement\_III},
	journal = {Journal of Epidemiology},
	author = {Nagai, Akiko and Hirata, Makoto and Kamatani, Yoichiro and Muto, Kaori and Matsuda, Koichi and Kiyohara, Yutaka and Ninomiya, Toshiharu and Tamakoshi, Akiko and Yamagata, Zentaro and Mushiroda, Taisei and Murakami, Yoshinori and Yuji, Koichiro and Furukawa, Yoichi and Zembutsu, Hitoshi and Tanaka, Toshihiro and Ohnishi, Yozo and Nakamura, Yusuke and {BioBank Japan Cooperative Hospital Group} and Kubo, Michiaki},
	year = {2017},
	pages = {S2--S8},
}

@article{rush_improving_2020,
	title = {Improving {Academic} {Biobank} {Value} and {Sustainability} {Through} an {Outputs} {Focus}},
	volume = {23},
	issn = {1098-3015},
	url = {https://www.sciencedirect.com/science/article/pii/S109830152032115X},
	doi = {10.1016/j.jval.2020.05.010},
	number = {8},
	urldate = {2024-04-09},
	journal = {Value in Health},
	author = {Rush, Amanda and Catchpoole, Daniel R. and Ling, Rod and Searles, Andrew and Watson, Peter H. and Byrne, Jennifer A.},
	month = aug,
	year = {2020},
	pages = {1072--1078},
}

@article{mabile_quantifying_2013,
	title = {Quantifying the use of bioresources for promoting their sharing in scientific research},
	volume = {2},
	issn = {2047-217X},
	url = {https://doi.org/10.1186/2047-217X-2-7},
	doi = {10.1186/2047-217X-2-7},
	number = {1},
	urldate = {2021-10-25},
	journal = {GigaScience},
	author = {Mabile, Laurence and Dalgleish, Raymond and Thorisson, Gudmundur A and Deschênes, Mylène and Hewitt, Robert and Carpenter, Jane and Bravo, Elena and Filocamo, Mirella and Gourraud, Pierre Antoine and Harris, Jennifer R and Hofman, Paul and Kauffmann, Francine and Muñoz-Fernàndez, Maria Angeles and Pasterk, Markus and Cambon-Thomsen, Anne and {BRIF working group}},
	month = dec,
	year = {2013},
}

@article{annaratone_basic_2021,
	title = {Basic principles of biobanking: from biological samples to precision medicine for patients},
	volume = {479},
	issn = {1432-2307},
	shorttitle = {Basic principles of biobanking},
	url = {https://doi.org/10.1007/s00428-021-03151-0},
	doi = {10.1007/s00428-021-03151-0},
	language = {en},
	number = {2},
	urldate = {2024-04-08},
	journal = {Virchows Archiv},
	author = {Annaratone, Laura and De Palma, Giuseppe and Bonizzi, Giuseppina and Sapino, Anna and Botti, Gerardo and Berrino, Enrico and Mannelli, Chiara and Arcella, Pamela and Di Martino, Simona and Steffan, Agostino and Daidone, Maria Grazia and Canzonieri, Vincenzo and Parodi, Barbara and Paradiso, Angelo Virgilio and Barberis, Massimo and Marchiò, Caterina and {On behalf of Alleanza Contro il Cancro (ACC) Pathology and Biobanking Working Group}},
	month = aug,
	year = {2021},
	pages = {233--246},
}

@article{vora_impacts_2015,
	title = {Impacts of a biobank: {Bridging} the gap in translational cancer medicine},
	volume = {36},
	copyright = {https://creativecommons.org/licenses/by-nc-nd/4.0/},
	issn = {0971-5851, 0975-2129},
	shorttitle = {Impacts of a biobank},
	url = {http://www.thieme-connect.de/DOI/DOI?10.4103/0971-5851.151773},
	doi = {10.4103/0971-5851.151773},
	language = {en},
	number = {01},
	urldate = {2024-04-08},
	journal = {Indian Journal of Medical and Paediatric Oncology},
	author = {Vora, Tushar and Thacker, Nirav},
	month = jan,
	year = {2015},
	pages = {17--23},
}

@article{swertz_towards_2022,
	title = {Towards an {Interoperable} {Ecosystem} of {Research} {Cohort} and {Real}-world {Data} {Catalogues} {Enabling} {Multi}-center {Studies}},
	volume = {31},
	copyright = {Georg Thieme Verlag KG Rüdigerstraße 14, 70469 Stuttgart, Germany},
	issn = {0943-4747, 2364-0502},
	url = {http://www.thieme-connect.de/DOI/DOI?10.1055/s-0042-1742522},
	doi = {10.1055/s-0042-1742522},
	language = {en},
	number = {1},
	urldate = {2024-04-04},
	journal = {Yearbook of Medical Informatics},
	author = {Swertz, Morris and Enckevort, Esther van and Oliveira, José Luis and Fortier, Isabel and Bergeron, Julie and Thurin, Nicolas H. and Hyde, Eleanor and Kellmann, Alexander and Pahoueshnja, Romin and Sturkenboom, Miriam and Cunnington, Marianne and Andersen, Anne-Marie Nybo and Marcon, Yannick and Gonçalves, Gonçalo and Gini, Rosa},
	month = aug,
	year = {2022},
	note = {Publisher: Georg Thieme Verlag KG},
	pages = {262--272},
}

@article{van_ommen_bbmri-eric_2015,
	title = {{BBMRI}-{ERIC} as a resource for pharmaceutical and life science industries: the development of biobank-based {Expert} {Centres}},
	volume = {23},
	copyright = {2015 The Author(s)},
	issn = {1476-5438},
	shorttitle = {{BBMRI}-{ERIC} as a resource for pharmaceutical and life science industries},
	url = {https://www.nature.com/articles/ejhg2014235},
	doi = {10.1038/ejhg.2014.235},
	language = {en},
	number = {7},
	urldate = {2024-04-03},
	journal = {European Journal of Human Genetics},
	author = {van Ommen, Gert-Jan B. and Törnwall, Outi and Bréchot, Christian and Dagher, Georges and Galli, Joakim and Hveem, Kristian and Landegren, Ulf and Luchinat, Claudio and Metspalu, Andres and Nilsson, Cecilia and Solesvik, Ove V. and Perola, Markus and Litton, Jan-Eric and Zatloukal, Kurt},
	month = jul,
	year = {2015},
	note = {Publisher: Nature Publishing Group},
	pages = {893--900},
}

@article{aksnes_citations_2019,
	title = {Citations, {Citation} {Indicators}, and {Research} {Quality}: {An} {Overview} of {Basic} {Concepts} and {Theories}},
	volume = {9},
	issn = {2158-2440},
	shorttitle = {Citations, {Citation} {Indicators}, and {Research} {Quality}},
	url = {https://doi.org/10.1177/2158244019829575},
	doi = {10.1177/2158244019829575},
	language = {en},
	number = {1},
	urldate = {2024-02-06},
	journal = {SAGE Open},
	author = {Aksnes, Dag W. and Langfeldt, Liv and Wouters, Paul},
	month = jan,
	year = {2019},
	note = {Publisher: SAGE Publications},
	pages = {2158244019829575},
}

@article{weis_learning_2021,
	title = {Learning on knowledge graph dynamics provides an early warning of impactful research},
	volume = {39},
	copyright = {2021 The Author(s), under exclusive licence to Springer Nature America, Inc.},
	issn = {1546-1696},
	url = {https://www.nature.com/articles/s41587-021-00907-6},
	doi = {10.1038/s41587-021-00907-6},
	language = {en},
	number = {10},
	urldate = {2024-02-01},
	journal = {Nature Biotechnology},
	author = {Weis, James W. and Jacobson, Joseph M.},
	month = oct,
	year = {2021},
	note = {Number: 10
Publisher: Nature Publishing Group},
	pages = {1300--1307},
}

@article{foster_tradition_2015,
	title = {Tradition and {Innovation} in {Scientists}’ {Research} {Strategies}},
	volume = {80},
	issn = {0003-1224},
	url = {https://doi.org/10.1177/0003122415601618},
	doi = {10.1177/0003122415601618},
	language = {en},
	number = {5},
	urldate = {2024-01-12},
	journal = {American Sociological Review},
	author = {Foster, Jacob G. and Rzhetsky, Andrey and Evans, James A.},
	month = oct,
	year = {2015},
	note = {Publisher: SAGE Publications Inc},
	pages = {875--908},
}

@book{wang_science_2021,
	title = {The {Science} of {Science}},
	isbn = {978-1-108-75825-3},
	language = {en},
	publisher = {Cambridge University Press},
	author = {Wang, Dashun and Barabási, Albert-László},
	month = mar,
	year = {2021},
	note = {Google-Books-ID: 3TIiEAAAQBAJ},
}

@article{gaziano_million_2016,
	title = {Million {Veteran} {Program}: {A} mega-biobank to study genetic influences on health and disease},
	volume = {70},
	issn = {0895-4356},
	shorttitle = {Million {Veteran} {Program}},
	url = {https://www.sciencedirect.com/science/article/pii/S0895435615004448},
	doi = {10.1016/j.jclinepi.2015.09.016},
	urldate = {2024-01-04},
	journal = {Journal of Clinical Epidemiology},
	author = {Gaziano, John Michael and Concato, John and Brophy, Mary and Fiore, Louis and Pyarajan, Saiju and Breeling, James and Whitbourne, Stacey and Deen, Jennifer and Shannon, Colleen and Humphries, Donald and Guarino, Peter and Aslan, Mihaela and Anderson, Daniel and LaFleur, Rene and Hammond, Timothy and Schaa, Kendra and Moser, Jennifer and Huang, Grant and Muralidhar, Sumitra and Przygodzki, Ronald and O'Leary, Timothy J.},
	month = feb,
	year = {2016},
	pages = {214--223},
}

@misc{meng_hidden_2023,
	title = {Hidden {Citations} {Obscure} {True} {Impact} in {Science}},
	url = {http://arxiv.org/abs/2310.16181},
	doi = {10.48550/arXiv.2310.16181},
	urldate = {2024-01-03},
	publisher = {arXiv},
	author = {Meng, Xiangyi and Varol, Onur and Barabási, Albert-László},
	month = oct,
	year = {2023},
	note = {arXiv:2310.16181 [physics]},
}

@article{hook_dimensions_2018,
	title = {Dimensions: {Building} {Context} for {Search} and {Evaluation}},
	volume = {3},
	issn = {2504-0537},
	shorttitle = {Dimensions},
	url = {https://www.frontiersin.org/articles/10.3389/frma.2018.00023},
	urldate = {2024-01-01},
	journal = {Frontiers in Research Metrics and Analytics},
	author = {Hook, Daniel W. and Porter, Simon J. and Herzog, Christian},
	year = {2018},
}

@article{bandrowski_decade_2022,
	title = {A decade of {GigaScience}: {What} can be learned from half a million {RRIDs} in the scientific literature?},
	volume = {11},
	issn = {2047-217X},
	shorttitle = {A decade of {GigaScience}},
	url = {https://doi.org/10.1093/gigascience/giac058},
	doi = {10.1093/gigascience/giac058},
	urldate = {2024-01-01},
	journal = {GigaScience},
	author = {Bandrowski, Anita},
	month = jan,
	year = {2022},
	pages = {giac058},
}

@article{fortunato_community_2010,
	title = {Community detection in graphs},
	volume = {486},
	issn = {0370-1573},
	url = {https://www.sciencedirect.com/science/article/pii/S0370157309002841},
	doi = {10.1016/j.physrep.2009.11.002},
	language = {en},
	number = {3},
	urldate = {2023-04-25},
	journal = {Physics Reports},
	author = {Fortunato, Santo},
	month = feb,
	year = {2010},
	pages = {75--174},
}

@article{fortunato_science_2018,
	title = {Science of science},
	volume = {359},
	url = {https://www.science.org/doi/full/10.1126/science.aao0185},
	doi = {10.1126/science.aao0185},
	number = {6379},
	urldate = {2023-01-27},
	journal = {Science},
	author = {Fortunato, Santo and Bergstrom, Carl T. and Börner, Katy and Evans, James A. and Helbing, Dirk and Milojević, Staša and Petersen, Alexander M. and Radicchi, Filippo and Sinatra, Roberta and Uzzi, Brian and Vespignani, Alessandro and Waltman, Ludo and Wang, Dashun and Barabási, Albert-László},
	month = mar,
	year = {2018},
	note = {Publisher: American Association for the Advancement of Science},
	pages = {eaao0185},
}

@article{shilo_axes_2020,
	title = {Axes of a revolution: challenges and promises of big data in healthcare},
	volume = {26},
	copyright = {2020 Springer Nature America, Inc.},
	issn = {1546-170X},
	shorttitle = {Axes of a revolution},
	url = {https://www.nature.com/articles/s41591-019-0727-5},
	doi = {10.1038/s41591-019-0727-5},
	language = {en},
	number = {1},
	urldate = {2022-11-25},
	journal = {Nature Medicine},
	author = {Shilo, Smadar and Rossman, Hagai and Segal, Eran},
	month = jan,
	year = {2020},
	note = {Number: 1
Publisher: Nature Publishing Group},
	pages = {29--38},
}

@article{illumina_inc_population_2020,
	title = {Population matters: {Biobanks} accelerate geno–pheno discoveries},
	copyright = {©2022 Macmillan Publishers Limited. All Rights Reserved.},
	shorttitle = {Population matters},
	url = {https://www.nature.com/articles/d42473-020-00238-1},
	language = {en},
	urldate = {2022-11-16},
	author = {Illumina Inc.},
	month = nov,
	year = {2020},
}

@article{lloyd-jones_framingham_2004,
	title = {Framingham risk score and prediction of lifetime risk for coronary heart disease},
	volume = {94},
	issn = {0002-9149},
	url = {https://www.sciencedirect.com/science/article/pii/S0002914904004370},
	doi = {10.1016/j.amjcard.2004.03.023},
	language = {en},
	number = {1},
	urldate = {2022-11-16},
	journal = {The American Journal of Cardiology},
	author = {Lloyd-Jones, Donald M and Wilson, Peter W. F and Larson, Martin G and Beiser, Alexa and Leip, Eric P and D'Agostino, Ralph B and Levy, Daniel},
	month = jul,
	year = {2004},
	pages = {20--24},
}

@article{mahmood_framingham_2014,
	title = {The {Framingham} {Heart} {Study} and the epidemiology of cardiovascular disease: a historical perspective},
	volume = {383},
	issn = {0140-6736},
	shorttitle = {The {Framingham} {Heart} {Study} and the epidemiology of cardiovascular disease},
	url = {https://www.sciencedirect.com/science/article/pii/S0140673613617523},
	doi = {10.1016/S0140-6736(13)61752-3},
	language = {en},
	number = {9921},
	urldate = {2022-11-16},
	journal = {The Lancet},
	author = {Mahmood, Syed S and Levy, Daniel and Vasan, Ramachandran S and Wang, Thomas J},
	month = mar,
	year = {2014},
	pages = {999--1008},
}

@article{griffith_collider_2020,
	title = {Collider bias undermines our understanding of {COVID}-19 disease risk and severity},
	volume = {11},
	copyright = {2020 The Author(s)},
	issn = {2041-1723},
	url = {https://www.nature.com/articles/s41467-020-19478-2},
	doi = {10.1038/s41467-020-19478-2},
	language = {en},
	number = {1},
	urldate = {2022-10-14},
	journal = {Nature Communications},
	author = {Griffith, Gareth J. and Morris, Tim T. and Tudball, Matthew J. and Herbert, Annie and Mancano, Giulia and Pike, Lindsey and Sharp, Gemma C. and Sterne, Jonathan and Palmer, Tom M. and Davey Smith, George and Tilling, Kate and Zuccolo, Luisa and Davies, Neil M. and Hemani, Gibran},
	month = nov,
	year = {2020},
	pages = {5749},
}

@article{ruth_using_2020,
	title = {Using human genetics to understand the disease impacts of testosterone in men and women},
	volume = {26},
	copyright = {2020 The Author(s), under exclusive licence to Springer Nature America, Inc.},
	issn = {1546-170X},
	url = {https://www.nature.com/articles/s41591-020-0751-5},
	doi = {10.1038/s41591-020-0751-5},
	language = {en},
	number = {2},
	urldate = {2022-10-14},
	journal = {Nature Medicine},
	author = {Ruth, Katherine S. and Day, Felix R. and Tyrrell, Jessica and Thompson, Deborah J. and Wood, Andrew R. and Mahajan, Anubha and Beaumont, Robin N. and Wittemans, Laura and Martin, Susan and Busch, Alexander S. and Erzurumluoglu, A. Mesut and Hollis, Benjamin and O’Mara, Tracy A. and McCarthy, Mark I. and Langenberg, Claudia and Easton, Douglas F. and Wareham, Nicholas J. and Burgess, Stephen and Murray, Anna and Ong, Ken K. and Frayling, Timothy M. and Perry, John R. B.},
	month = feb,
	year = {2020},
	pages = {252--258},
}

@article{turro_whole-genome_2020,
	title = {Whole-genome sequencing of patients with rare diseases in a national health system},
	volume = {583},
	copyright = {2020 The Author(s), under exclusive licence to Springer Nature Limited},
	issn = {1476-4687},
	url = {https://www.nature.com/articles/s41586-020-2434-2},
	doi = {10.1038/s41586-020-2434-2},
	language = {en},
	number = {7814},
	urldate = {2022-10-14},
	journal = {Nature},
	author = {Turro, Ernest and Astle, William J. and Megy, Karyn and Gräf, Stefan and Greene, Daniel and Shamardina, Olga and Allen, Hana Lango and Sanchis-Juan, Alba and Frontini, Mattia and Thys, Chantal and Stephens, Jonathan and Mapeta, Rutendo and Burren, Oliver S. and Downes, Kate and Haimel, Matthias and Tuna, Salih and Deevi, Sri V. V. and Aitman, Timothy J. and Bennett, David L. and Calleja, Paul and Carss, Keren and Caulfield, Mark J. and Chinnery, Patrick F. and Dixon, Peter H. and Gale, Daniel P. and James, Roger and Koziell, Ania and Laffan, Michael A. and Levine, Adam P. and Maher, Eamonn R. and Markus, Hugh S. and Morales, Joannella and Morrell, Nicholas W. and Mumford, Andrew D. and Ormondroyd, Elizabeth and Rankin, Stuart and Rendon, Augusto and Richardson, Sylvia and Roberts, Irene and Roy, Noemi B. A. and Saleem, Moin A. and Smith, Kenneth G. C. and Stark, Hannah and Tan, Rhea Y. Y. and Themistocleous, Andreas C. and Thrasher, Adrian J. and Watkins, Hugh and Webster, Andrew R. and Wilkins, Martin R. and Williamson, Catherine and Whitworth, James and Humphray, Sean and Bentley, David R. and Kingston, Nathalie and Walker, Neil and Bradley, John R. and Ashford, Sofie and Penkett, Christopher J. and Freson, Kathleen and Stirrups, Kathleen E. and Raymond, F. Lucy and Ouwehand, Willem H.},
	month = jul,
	year = {2020},
	pages = {96--102},
}

@article{warrington_maternal_2019,
	title = {Maternal and fetal genetic effects on birth weight and their relevance to cardio-metabolic risk factors},
	volume = {51},
	copyright = {2019 The Author(s), under exclusive licence to Springer Nature America, Inc.},
	issn = {1546-1718},
	url = {https://www.nature.com/articles/s41588-019-0403-1},
	doi = {10.1038/s41588-019-0403-1},
	language = {en},
	number = {5},
	urldate = {2022-10-14},
	journal = {Nature Genetics},
	author = {Warrington, Nicole M. and Beaumont, Robin N. and Horikoshi, Momoko and Day, Felix R. and Helgeland, Øyvind and Laurin, Charles and Bacelis, Jonas and Peng, Shouneng and Hao, Ke and Feenstra, Bjarke and Wood, Andrew R. and Mahajan, Anubha and Tyrrell, Jessica and Robertson, Neil R. and Rayner, N. William and Qiao, Zhen and Moen, Gunn-Helen and Vaudel, Marc and Marsit, Carmen J. and Chen, Jia and Nodzenski, Michael and Schnurr, Theresia M. and Zafarmand, Mohammad H. and Bradfield, Jonathan P. and Grarup, Niels and Kooijman, Marjolein N. and Li-Gao, Ruifang and Geller, Frank and Ahluwalia, Tarunveer S. and Paternoster, Lavinia and Rueedi, Rico and Huikari, Ville and Hottenga, Jouke-Jan and Lyytikäinen, Leo-Pekka and Cavadino, Alana and Metrustry, Sarah and Cousminer, Diana L. and Wu, Ying and Thiering, Elisabeth and Wang, Carol A. and Have, Christian T. and Vilor-Tejedor, Natalia and Joshi, Peter K. and Painter, Jodie N. and Ntalla, Ioanna and Myhre, Ronny and Pitkänen, Niina and van Leeuwen, Elisabeth M. and Joro, Raimo and Lagou, Vasiliki and Richmond, Rebecca C. and Espinosa, Ana and Barton, Sheila J. and Inskip, Hazel M. and Holloway, John W. and Santa-Marina, Loreto and Estivill, Xavier and Ang, Wei and Marsh, Julie A. and Reichetzeder, Christoph and Marullo, Letizia and Hocher, Berthold and Lunetta, Kathryn L. and Murabito, Joanne M. and Relton, Caroline L. and Kogevinas, Manolis and Chatzi, Leda and Allard, Catherine and Bouchard, Luigi and Hivert, Marie-France and Zhang, Ge and Muglia, Louis J. and Heikkinen, Jani and Morgen, Camilla S. and van Kampen, Antoine H. C. and van Schaik, Barbera D. C. and Mentch, Frank D. and Langenberg, Claudia and Luan, Jian’an and Scott, Robert A. and Zhao, Jing Hua and Hemani, Gibran and Ring, Susan M. and Bennett, Amanda J. and Gaulton, Kyle J. and Fernandez-Tajes, Juan and van Zuydam, Natalie R. and Medina-Gomez, Carolina and de Haan, Hugoline G. and Rosendaal, Frits R. and Kutalik, Zoltán and Marques-Vidal, Pedro and Das, Shikta and Willemsen, Gonneke and Mbarek, Hamdi and Müller-Nurasyid, Martina and Standl, Marie and Appel, Emil V. R. and Fonvig, Cilius E. and Trier, Caecilie and van Beijsterveldt, Catharina E. M. and Murcia, Mario and Bustamante, Mariona and Bonas-Guarch, Sílvia and Hougaard, David M. and Mercader, Josep M. and Linneberg, Allan and Schraut, Katharina E. and Lind, Penelope A. and Medland, Sarah E. and Shields, Beverley M. and Knight, Bridget A. and Chai, Jin-Fang and Panoutsopoulou, Kalliope and Bartels, Meike and Sánchez, Friman and Stokholm, Jakob and Torrents, David and Vinding, Rebecca K. and Willems, Sara M. and Atalay, Mustafa and Chawes, Bo L. and Kovacs, Peter and Prokopenko, Inga and Tuke, Marcus A. and Yaghootkar, Hanieh and Ruth, Katherine S. and Jones, Samuel E. and Loh, Po-Ru and Murray, Anna and Weedon, Michael N. and Tönjes, Anke and Stumvoll, Michael and Michaelsen, Kim F. and Eloranta, Aino-Maija and Lakka, Timo A. and van Duijn, Cornelia M. and Kiess, Wieland and Körner, Antje and Niinikoski, Harri and Pahkala, Katja and Raitakari, Olli T. and Jacobsson, Bo and Zeggini, Eleftheria and Dedoussis, George V. and Teo, Yik-Ying and Saw, Seang-Mei and Montgomery, Grant W. and Campbell, Harry and Wilson, James F. and Vrijkotte, Tanja G. M. and Vrijheid, Martine and de Geus, Eco J. C. N. and Hayes, M. Geoffrey and Kadarmideen, Haja N. and Holm, Jens-Christian and Beilin, Lawrence J. and Pennell, Craig E. and Heinrich, Joachim and Adair, Linda S. and Borja, Judith B. and Mohlke, Karen L. and Eriksson, Johan G. and Widén, Elisabeth E. and Hattersley, Andrew T. and Spector, Tim D. and Kähönen, Mika and Viikari, Jorma S. and Lehtimäki, Terho and Boomsma, Dorret I. and Sebert, Sylvain and Vollenweider, Peter and Sørensen, Thorkild I. A. and Bisgaard, Hans and Bønnelykke, Klaus and Murray, Jeffrey C. and Melbye, Mads and Nohr, Ellen A. and Mook-Kanamori, Dennis O. and Rivadeneira, Fernando and Hofman, Albert and Felix, Janine F. and Jaddoe, Vincent W. V. and Hansen, Torben and Pisinger, Charlotta and Vaag, Allan A. and Pedersen, Oluf and Uitterlinden, André G. and Järvelin, Marjo-Riitta and Power, Christine and Hyppönen, Elina and Scholtens, Denise M. and Lowe, William L. and Davey Smith, George and Timpson, Nicholas J. and Morris, Andrew P. and Wareham, Nicholas J. and Hakonarson, Hakon and Grant, Struan F. A. and Frayling, Timothy M. and Lawlor, Debbie A. and Njølstad, Pål R. and Johansson, Stefan and Ong, Ken K. and McCarthy, Mark I. and Perry, John R. B. and Evans, David M. and Freathy, Rachel M.},
	month = may,
	year = {2019},
	pages = {804--814},
}

@article{gates_wealth_2021,
	title = {A wealth of discovery built on the {Human} {Genome} {Project} — by the numbers},
	volume = {590},
	copyright = {2021 Nature},
	url = {https://www.nature.com/articles/d41586-021-00314-6},
	doi = {10.1038/d41586-021-00314-6},
	language = {en},
	number = {7845},
	urldate = {2022-09-07},
	journal = {Nature},
	author = {Gates, Alexander J. and Gysi, Deisy Morselli and Kellis, Manolis and Barabási, Albert-László},
	month = feb,
	year = {2021},
	pages = {212--215},
}

@article{park_10_2009,
	title = {10 {Ideas} {Changing} the {World} {Right} {Now} - {TIME}},
	issn = {0040-781X},
	url = {https://content.time.com/time/specials/packages/article/0,28804,1884779_1884782_1884766,00.html},
	language = {en-US},
	urldate = {2022-08-15},
	journal = {Time},
	author = {Park, Alice},
	month = mar,
	year = {2009},
}

@article{shen_collective_2014,
	title = {Collective credit allocation in science},
	volume = {111},
	url = {https://www.pnas.org/doi/full/10.1073/pnas.1401992111},
	doi = {10.1073/pnas.1401992111},
	number = {34},
	urldate = {2022-07-01},
	journal = {Proceedings of the National Academy of Sciences},
	author = {Shen, Hua-Wei and Barabási, Albert-László},
	month = aug,
	year = {2014},
	pages = {12325--12330},
}

@article{simonton_scientific_2013,
	title = {Scientific genius is extinct},
	volume = {493},
	copyright = {2013 Nature Publishing Group, a division of Macmillan Publishers Limited. All Rights Reserved.},
	issn = {1476-4687},
	url = {https://www.nature.com/articles/493602a},
	doi = {10.1038/493602a},
	language = {en},
	number = {7434},
	urldate = {2022-07-01},
	journal = {Nature},
	author = {Simonton, Dean Keith},
	month = jan,
	year = {2013},
	pages = {602--602},
}

@article{serwadda_open_2018,
	title = {Open data sharing and the {Global} {South}—{Who} benefits?},
	volume = {359},
	url = {https://www.science.org/doi/full/10.1126/science.aap8395},
	doi = {10.1126/science.aap8395},
	number = {6376},
	urldate = {2022-07-01},
	journal = {Science},
	author = {Serwadda, David and Ndebele, Paul and Grabowski, M. Kate and Bajunirwe, Francis and Wanyenze, Rhoda K.},
	month = feb,
	year = {2018},
	pages = {642--643},
}

@book{hollington_concepts_2015,
	address = {Cologne, Germany},
	title = {Concepts of the {Global} {South}},
	volume = {2015/1},
	url = {http://www.uni-koeln.de/},
	language = {de},
	urldate = {2022-06-29},
	publisher = {Global South Studies Center Cologne},
	author = {Hollington, Andrea and Salverda, Tijo and Schwarz, Tobias and Tappe, Oliver},
	editor = {Hollington, Andrea and Salverda, Tijo and Schwarz, Tobias and Tappe, Oliver},
	collaborator = {Damsa-Ard, Regina and Weck, Frederik and Rath, Christine},
	month = jan,
	year = {2015},
}

@book{development_health_1990,
	title = {Health {Research}: {Essential} {Link} to {Equity} in {Development}},
	isbn = {9780195208382},
	shorttitle = {Health {Research}},
	language = {en},
	publisher = {Oxford University Press},
	author = {Development, Commission on Health Research for},
	year = {1990},
	note = {Google-Books-ID: gY0iovWA8voC},
}

@article{haelewaters_ten_2021,
	title = {Ten simple rules for {Global} {North} researchers to stop perpetuating helicopter research in the {Global} {South}},
	volume = {17},
	issn = {1553-734X},
	url = {https://www.ncbi.nlm.nih.gov/pmc/articles/PMC8376010/},
	doi = {10.1371/journal.pcbi.1009277},
	number = {8},
	urldate = {2022-06-15},
	journal = {PLoS Computational Biology},
	author = {Haelewaters, Danny and Hofmann, Tina A. and Romero-Olivares, Adriana L.},
	month = aug,
	year = {2021},
	pmid = {34411090},
	pmcid = {PMC8376010},
	pages = {e1009277},
}

@article{mills_scientometric_2019,
	title = {A scientometric review of genome-wide association studies},
	volume = {2},
	copyright = {2019 The Author(s)},
	issn = {2399-3642},
	url = {https://www.nature.com/articles/s42003-018-0261-x},
	doi = {10.1038/s42003-018-0261-x},
	language = {en},
	number = {1},
	urldate = {2022-06-12},
	journal = {Communications Biology},
	author = {Mills, Melinda C. and Rahal, Charles},
	month = jan,
	year = {2019},
	note = {Number: 1
Publisher: Nature Publishing Group},
	pages = {1--11},
}

@article{blei_latent_2003,
	title = {Latent {Dirichlet} {Allocation}},
	volume = {3},
	issn = {ISSN 1533-7928},
	url = {https://www.jmlr.org/papers/v3/blei03a.html},
	number = {Jan},
	urldate = {2022-06-02},
	journal = {Journal of Machine Learning Research},
	author = {Blei, David M. and Ng, Andrew Y. and Jordan, Michael I.},
	year = {2003},
	pages = {993--1022},
}

@article{lawson_is_2020,
	title = {Is population structure in the genetic biobank era irrelevant, a challenge, or an opportunity?},
	volume = {139},
	issn = {1432-1203},
	url = {https://doi.org/10.1007/s00439-019-02014-8},
	doi = {10.1007/s00439-019-02014-8},
	language = {en},
	number = {1},
	urldate = {2022-05-30},
	journal = {Human Genetics},
	author = {Lawson, Daniel John and Davies, Neil Martin and Haworth, Simon and Ashraf, Bilal and Howe, Laurence and Crawford, Andrew and Hemani, Gibran and Davey Smith, George and Timpson, Nicholas John},
	month = jan,
	year = {2020},
	pages = {23--41},
}

@article{pingault_using_2018,
	title = {Using genetic data to strengthen causal inference in observational research},
	volume = {19},
	copyright = {2018 Macmillan Publishers Ltd., part of Springer Nature},
	issn = {1471-0064},
	url = {https://www.nature.com/articles/s41576-018-0020-3},
	doi = {10.1038/s41576-018-0020-3},
	language = {en},
	number = {9},
	urldate = {2022-05-30},
	journal = {Nature Reviews Genetics},
	author = {Pingault, Jean-Baptiste and O’Reilly, Paul F. and Schoeler, Tabea and Ploubidis, George B. and Rijsdijk, Frühling and Dudbridge, Frank},
	month = sep,
	year = {2018},
	note = {Number: 9
Publisher: Nature Publishing Group},
	pages = {566--580},
}

@article{gaskell_biobanks_2011,
	title = {Biobanks need publicity},
	volume = {471},
	copyright = {2011 Nature Publishing Group, a division of Macmillan Publishers Limited. All Rights Reserved.},
	issn = {1476-4687},
	url = {https://www.nature.com/articles/471159a},
	doi = {10.1038/471159a},
	language = {en},
	number = {7337},
	urldate = {2022-05-27},
	journal = {Nature},
	author = {Gaskell, George and Gottweis, Herbert},
	month = mar,
	year = {2011},
	note = {Number: 7337
Publisher: Nature Publishing Group},
	pages = {159--160},
}

@article{wichmann_comprehensive_2011,
	title = {Comprehensive catalog of {European} biobanks},
	volume = {29},
	copyright = {2011 Nature Publishing Group, a division of Macmillan Publishers Limited. All Rights Reserved.},
	issn = {1546-1696},
	url = {https://www.nature.com/articles/nbt.1958},
	doi = {10.1038/nbt.1958},
	language = {en},
	number = {9},
	urldate = {2022-05-27},
	journal = {Nature Biotechnology},
	author = {Wichmann, H.-Erich and Kuhn, Klaus A. and Waldenberger, Melanie and Schmelcher, Dominik and Schuffenhauer, Simone and Meitinger, Thomas and Wurst, Sebastian H. R. and Lamla, Gregor and Fortier, Isabel and Burton, Paul R. and Peltonen, Leena and Perola, Markus and Metspalu, Andres and Riegman, Peter and Landegren, Ulf and Taussig, Michael J. and Litton, Jan-Eric and Fransson, Martin N. and Eder, Johann and Cambon-Thomsen, Anne and Bovenberg, Jasper and Dagher, Georges and van Ommen, Gert-Jan and Griffith, Michael and Yuille, Martin and Zatloukal, Kurt},
	month = sep,
	year = {2011},
	note = {Number: 9
Publisher: Nature Publishing Group},
	pages = {795--797},
}

@article{fried_cardiovascular_1991,
	title = {The cardiovascular health study: {Design} and rationale},
	volume = {1},
	issn = {1047-2797},
	shorttitle = {The cardiovascular health study},
	url = {https://www.sciencedirect.com/science/article/pii/104727979190005W},
	doi = {10.1016/1047-2797(91)90005-W},
	language = {en},
	number = {3},
	urldate = {2022-04-13},
	journal = {Annals of Epidemiology},
	author = {Fried, Linda P. and Borhani, Nemat O. and Enright, Paul and Furberg, Curt D. and Gardin, Julius M. and Kronmal, Richard A. and Kuller, Lewis H. and Manolio, Teri A. and Mittelmark, Maurice B. and Newman, Anne and O'Leary, Daniel H. and Psaty, Bruce and Rautaharju, Pentti and Tracy, Russell P. and Weiler, Philip G.},
	month = feb,
	year = {1991},
	pages = {263--276},
}

@article{lawrence_barriers_2020,
	title = {The {Barriers} and {Motivators} to {Using} {Human} {Tissues} for {Research}: {The} {Views} of {UK}-{Based} {Biomedical} {Researchers}},
	volume = {18},
	issn = {1947-5535},
	shorttitle = {The {Barriers} and {Motivators} to {Using} {Human} {Tissues} for {Research}},
	url = {https://www.liebertpub.com/doi/full/10.1089/bio.2019.0138},
	doi = {10.1089/bio.2019.0138},
	number = {4},
	urldate = {2022-03-09},
	journal = {Biopreservation and Biobanking},
	author = {Lawrence, Emma and Sims, Jessica and Gander, Amir and Garibaldi, Jonathan M. and Fuller, Barry and Davidson, Brian and Quinlan, Philip R.},
	month = aug,
	year = {2020},
	note = {Publisher: Mary Ann Liebert, Inc., publishers},
	pages = {266--273},
}

@article{kannel_factors_1961,
	title = {Factors of {Risk} in the {Development} of {Coronary} {Heart} {Disease}—{Six}-{Year} {Follow}-up {Experience}},
	volume = {55},
	issn = {0003-4819},
	url = {https://www.acpjournals.org/doi/abs/10.7326/0003-4819-55-1-33},
	doi = {10.7326/0003-4819-55-1-33},
	number = {1},
	urldate = {2022-02-14},
	journal = {Annals of Internal Medicine},
	author = {Kannel, William B. and Dawber, Thomas R. and Kagan, Abraham and Revotskie, Nicholas and Stokes, Joseph},
	month = jul,
	year = {1961},
	note = {Publisher: American College of Physicians},
	pages = {33--50},
}

@article{caulfield_review_2014,
	title = {A review of the key issues associated with the commercialization of biobanks},
	volume = {1},
	issn = {2053-9711},
	url = {https://doi.org/10.1093/jlb/lst004},
	doi = {10.1093/jlb/lst004},
	number = {1},
	urldate = {2022-02-14},
	journal = {Journal of Law and the Biosciences},
	author = {Caulfield, Timothy and Burningham, Sarah and Joly, Yann and Master, Zubin and Shabani, Mahsa and Borry, Pascal and Becker, Allan and Burgess, Michael and Calder, Kathryn and Critchley, Christine and Edwards, Kelly and Fullerton, Stephanie M. and Gottweis, Herbert and Hyde-Lay, Robyn and Illes, Judy and Isasi, Rosario and Kato, Kazuto and Kaye, Jane and Knoppers, Bartha and Lynch, John and McGuire, Amy and Meslin, Eric and Nicol, Dianne and O’Doherty, Kieran and Ogbogu, Ubaka and Otlowski, Margaret and Pullman, Daryl and Ries, Nola and Scott, Chris and Sears, Malcolm and Wallace, Helen and Zawati, Ma'n H.},
	month = mar,
	year = {2014},
	pages = {94--110},
}

@article{norman_cohort_2009,
	title = {Cohort {Profile}: {The} {Health} {In} {Men} {Study} ({HIMS})},
	volume = {38},
	issn = {0300-5771},
	shorttitle = {Cohort {Profile}},
	url = {https://doi.org/10.1093/ije/dyn041},
	doi = {10.1093/ije/dyn041},
	number = {1},
	urldate = {2022-01-19},
	journal = {International Journal of Epidemiology},
	author = {Norman, Paul E and Flicker, Leon and Almeida, Osvaldo P and Hankey, Graeme J and Hyde, Zoë and Jamrozik, Konrad},
	month = feb,
	year = {2009},
	pages = {48--52},
}

@article{carlson_maine_1994,
	title = {The {Maine} {Women}'s {Health} {Study}: {I}. {Outcomes} of hysterectomy},
	volume = {83},
	issn = {1873-233X},
	shorttitle = {The {Maine} {Women}'s {Health} {Study}},
	url = {https://doi.org/10.1097/00006250-199404000-00012},
	doi = {10.1097/00006250-199404000-00012},
	language = {eng},
	number = {4},
	urldate = {2022-01-19},
	journal = {Obstetrics and gynecology},
	author = {Carlson, K J and Miller, B A and Fowler, F J},
	month = apr,
	year = {1994},
	pmid = {8134066},
	pages = {556--565},
}

@article{arnold_cohort_2019,
	title = {Cohort profile: the {Australian} {Longitudinal} {Study} of {Adults} with {Autism} ({ALSAA})},
	volume = {9},
	copyright = {© Author(s) (or their employer(s)) 2019. Re-use permitted under CC BY-NC. No commercial re-use. See rights and permissions. Published by BMJ.. This is an open access article distributed in accordance with the Creative Commons Attribution Non Commercial (CC BY-NC 4.0) license, which permits others to distribute, remix, adapt, build upon this work non-commercially, and license their derivative works on different terms, provided the original work is properly cited, appropriate credit is given, any changes made indicated, and the use is non-commercial. See: http://creativecommons.org/licenses/by-nc/4.0/.},
	issn = {2044-6055, 2044-6055},
	shorttitle = {Cohort profile},
	url = {https://bmjopen.bmj.com/content/9/12/e030798},
	doi = {10.1136/bmjopen-2019-030798},
	language = {en},
	number = {12},
	urldate = {2022-01-19},
	journal = {BMJ Open},
	author = {Arnold, Samuel and Foley, Kitty-Rose and Hwang, Ye In (Jane) and Richdale, Amanda L. and Uljarevic, Mirko and Lawson, Lauren P. and Cai, Ru Ying and Falkmer, Torbjorn and Falkmer, Marita and Lennox, Nick G. and Urbanowicz, Anna and Trollor, Julian},
	month = dec,
	year = {2019},
	pmid = {31806608},
	note = {Publisher: British Medical Journal Publishing Group
Section: Mental health},
	pages = {e030798},
}

@article{repka_cataract_2016,
	title = {Cataract {Surgery} in {Children} from {Birth} to {Less} than 13 {Years} of {Age}: {Baseline} {Characteristics} of the {Cohort}},
	volume = {123},
	issn = {0161-6420},
	shorttitle = {Cataract {Surgery} in {Children} from {Birth} to {Less} than 13 {Years} of {Age}},
	url = {https://www.sciencedirect.com/science/article/pii/S0161642016311654},
	doi = {10.1016/j.ophtha.2016.09.003},
	language = {en},
	number = {12},
	urldate = {2022-01-19},
	journal = {Ophthalmology},
	author = {Repka, Michael X. and Dean, Trevano W. and Lazar, Elizabeth L. and Yen, Kimberly G. and Lenhart, Phoebe D. and Freedman, Sharon F. and Hug, Denise and Rahmani, Bahram and Wang, Serena X. and Kraker, Raymond T. and Wallace, David K.},
	month = dec,
	year = {2016},
	pages = {2462--2473},
}

@article{marbury_indoor_1996,
	title = {The {Indoor} {Air} and {Children}'s {Health} {Study}: {Methods} and {Incidence} {Rates}},
	volume = {7},
	issn = {1044-3983},
	shorttitle = {The {Indoor} {Air} and {Children}'s {Health} {Study}},
	url = {https://www.jstor.org/stable/3703031},
	number = {2},
	urldate = {2022-01-19},
	journal = {Epidemiology},
	author = {Marbury, Marian C. and Maldonado, George and Waller, Lance},
	year = {1996},
	note = {Publisher: Lippincott Williams \& Wilkins},
	pages = {166--174},
}

@article{otoole_australian_1996,
	title = {The {Australian} {Vietnam} {Veterans} {Health} {Study}: {III}. {Psychological} {Health} of {Australian} {Vietnam} {Veterans} and its {Relationship} to {Combat}},
	volume = {25},
	issn = {0300-5771},
	shorttitle = {The {Australian} {Vietnam} {Veterans} {Health} {Study}},
	url = {https://doi.org/10.1093/ije/25.2.331},
	doi = {10.1093/ije/25.2.331},
	number = {2},
	urldate = {2022-01-19},
	journal = {International Journal of Epidemiology},
	author = {O'TOOLE, BRIAN I and MARSHALL, RICHARD P and GRAYSON, DAVID A and SCHURECK, RALPH J and DOBSON, MATTHEW and FFRENCH, MARGOT and PULVERTAFT, BELINDA and MELDRUM, LENORE and BOLTON, JAMES and VENNARD, JULIENNE},
	month = apr,
	year = {1996},
	pages = {331--340},
}

@article{sachdev_sydney_2010,
	title = {The {Sydney} {Memory} and {Ageing} {Study} ({MAS}): methodology and baseline medical and neuropsychiatric characteristics of an elderly epidemiological non-demented cohort of {Australians} aged 70–90 years},
	volume = {22},
	issn = {1741-203X, 1041-6102},
	shorttitle = {The {Sydney} {Memory} and {Ageing} {Study} ({MAS})},
	url = {https://www.cambridge.org/core/journals/international-psychogeriatrics/article/sydney-memory-and-ageing-study-mas-methodology-and-baseline-medical-and-neuropsychiatric-characteristics-of-an-elderly-epidemiological-nondemented-cohort-of-australians-aged-7090-years/597631961DCE67C34CE595A549D7A920},
	doi = {10.1017/S1041610210001067},
	language = {en},
	number = {8},
	urldate = {2022-01-19},
	journal = {International Psychogeriatrics},
	author = {Sachdev, Perminder S. and Brodaty, Henry and Reppermund, Simone and Kochan, Nicole A. and Trollor, Julian N. and Draper, Brian and Slavin, Melissa J. and Crawford, John and Kang, Kristan and Broe, G. Anthony and Mather, Karen A. and Lux, Ora and Team, the Memory {and} Ageing Study},
	month = dec,
	year = {2010},
	note = {Publisher: Cambridge University Press},
	pages = {1248--1264},
}

@article{copeland_characteristics_2011,
	title = {Characteristics of {Adolescents} and {Youth} with {Recent}-{Onset} {Type} 2 {Diabetes}: {The} {TODAY} {Cohort} at {Baseline}},
	volume = {96},
	issn = {0021-972X},
	shorttitle = {Characteristics of {Adolescents} and {Youth} with {Recent}-{Onset} {Type} 2 {Diabetes}},
	url = {https://doi.org/10.1210/jc.2010-1642},
	doi = {10.1210/jc.2010-1642},
	number = {1},
	urldate = {2022-01-19},
	journal = {The Journal of Clinical Endocrinology \& Metabolism},
	author = {Copeland, Kenneth C. and Zeitler, Philip and Geffner, Mitchell and Guandalini, Cindy and Higgins, Janine and Hirst, Kathryn and Kaufman, Francine R. and Linder, Barbara and Marcovina, Santica and McGuigan, Paul and Pyle, Laura and Tamborlane, William and Willi, Steven},
	month = jan,
	year = {2011},
	pages = {159--167},
}

@article{moffet_cohort_2009,
	title = {Cohort {Profile}: {The} {Diabetes} {Study} of {Northern} {California} ({DISTANCE})—objectives and design of a survey follow-up study of social health disparities in a managed care population†},
	volume = {38},
	issn = {0300-5771},
	shorttitle = {Cohort {Profile}},
	url = {https://doi.org/10.1093/ije/dyn040},
	doi = {10.1093/ije/dyn040},
	number = {1},
	urldate = {2022-01-19},
	journal = {International Journal of Epidemiology},
	author = {Moffet, Howard H and Adler, Nancy and Schillinger, Dean and Ahmed, Ameena T and Laraia, Barbara and Selby, Joe V and Neugebauer, Romain and Liu, Jennifer Y and Parker, Melissa M and Warton, Margaret and Karter, Andrew J},
	month = feb,
	year = {2009},
	pages = {38--47},
}

@article{the_swiss_hiv_cohort_study_cohort_2010,
	title = {Cohort {Profile}: {The} {Swiss} {HIV} {Cohort} {Study}},
	volume = {39},
	issn = {0300-5771},
	shorttitle = {Cohort {Profile}},
	url = {https://doi.org/10.1093/ije/dyp321},
	doi = {10.1093/ije/dyp321},
	number = {5},
	urldate = {2022-01-19},
	journal = {International Journal of Epidemiology},
	author = {{The Swiss HIV Cohort Study} and Schoeni-Affolter, Franziska and Ledergerber, Bruno and Rickenbach, Martin and Rudin, Christoph and Günthard, Huldrych F. and Telenti, Amalio and Furrer, Hansjakob and Yerly, Sabine and Francioli, Patrick},
	month = oct,
	year = {2010},
	pages = {1179--1189},
}

@article{tanser_cohort_2008,
	title = {Cohort {Profile}: {Africa} {Centre} {Demographic} {Information} {System} ({ACDIS}) and population-based {HIV} survey},
	volume = {37},
	issn = {0300-5771},
	shorttitle = {Cohort {Profile}},
	url = {https://doi.org/10.1093/ije/dym211},
	doi = {10.1093/ije/dym211},
	number = {5},
	urldate = {2022-01-19},
	journal = {International Journal of Epidemiology},
	author = {Tanser, Frank and Hosegood, Victoria and Bärnighausen, Till and Herbst, Kobus and Nyirenda, Makandwe and Muhwava, William and Newell, Colin and Viljoen, Johannes and Mutevedzi, Tinofa and Newell, Marie-Louise},
	month = oct,
	year = {2008},
	pages = {956--962},
}

@article{bennett_rush_2005,
	title = {The {Rush} {Memory} and {Aging} {Project}: {Study} {Design} and {Baseline} {Characteristics} of the {Study} {Cohort}},
	volume = {25},
	issn = {0251-5350, 1423-0208},
	shorttitle = {The {Rush} {Memory} and {Aging} {Project}},
	url = {https://www.karger.com/Article/FullText/87446},
	doi = {10.1159/000087446},
	language = {english},
	number = {4},
	urldate = {2022-01-19},
	journal = {Neuroepidemiology},
	author = {Bennett, David A. and Schneider, Julie A. and Buchman, Aron S. and Leon, Carlos Mendes de and Bienias, Julia L. and Wilson, Robert S.},
	year = {2005},
	pmid = {16103727},
	note = {Publisher: Karger Publishers},
	pages = {163--175},
}

@article{odonoghue_how_2021,
	title = {How {Many} {Health} {Research} {Biobanks} {Are} {There}?},
	issn = {1947-5535},
	url = {https://www.liebertpub.com/doi/full/10.1089/bio.2021.0063},
	doi = {10.1089/bio.2021.0063},
	urldate = {2022-01-14},
	journal = {Biopreservation and Biobanking},
	author = {O'Donoghue, Sheila and Dee, Simon and Byrne, Jennifer A. and Watson, Peter Hamilton},
	month = sep,
	year = {2021},
	note = {Publisher: Mary Ann Liebert, Inc., publishers},
}

@article{kleiderman_author_2018,
	title = {The author who wasn’t there? {Fairness} and attribution in publications following access to population biobanks},
	volume = {13},
	issn = {1932-6203},
	shorttitle = {The author who wasn’t there?},
	url = {https://journals.plos.org/plosone/article?id=10.1371/journal.pone.0194997},
	doi = {10.1371/journal.pone.0194997},
	language = {en},
	number = {3},
	urldate = {2022-01-14},
	journal = {PLOS ONE},
	author = {Kleiderman, Erika and Pack, Amy and Borry, Pascal and Zawati, Ma’n},
	month = mar,
	year = {2018},
	note = {Publisher: Public Library of Science},
	pages = {e0194997},
}

@article{chen_china_2011,
	title = {China {Kadoorie} {Biobank} of 0.5 million people: survey methods, baseline characteristics and long-term follow-up},
	volume = {40},
	issn = {0300-5771},
	shorttitle = {China {Kadoorie} {Biobank} of 0.5 million people},
	url = {https://doi.org/10.1093/ije/dyr120},
	doi = {10.1093/ije/dyr120},
	number = {6},
	urldate = {2022-01-12},
	journal = {International Journal of Epidemiology},
	author = {Chen, Zhengming and Chen, Junshi and Collins, Rory and Guo, Yu and Peto, Richard and Wu, Fan and Li, Liming and {on behalf of the China Kadoorie Biobank (CKB) collaborative group}},
	month = dec,
	year = {2011},
	pages = {1652--1666},
}

@article{krokstad_cohort_2013,
	title = {Cohort {Profile}: {The} {HUNT} {Study}, {Norway}},
	volume = {42},
	issn = {0300-5771},
	shorttitle = {Cohort {Profile}},
	url = {https://doi.org/10.1093/ije/dys095},
	doi = {10.1093/ije/dys095},
	number = {4},
	urldate = {2022-01-12},
	journal = {International Journal of Epidemiology},
	author = {Krokstad, S and Langhammer, A and Hveem, K and Holmen, TL and Midthjell, K and Stene, TR and Bratberg, G and Heggland, J and Holmen, J},
	month = aug,
	year = {2013},
	pages = {968--977},
}

@article{hajar_framingham_2016,
	title = {Framingham {Contribution} to {Cardiovascular} {Disease}},
	volume = {17},
	issn = {1995-705X},
	url = {https://www.ncbi.nlm.nih.gov/pmc/articles/PMC4966216/},
	doi = {10.4103/1995-705X.185130},
	number = {2},
	urldate = {2022-01-07},
	journal = {Heart Views : The Official Journal of the Gulf Heart Association},
	author = {Hajar, Rachel},
	year = {2016},
	pmid = {27512540},
	pmcid = {PMC4966216},
	pages = {78--81},
}

@article{kinkorova_biobanks_2018,
	title = {Biobanks in {Horizon} 2020: sustainability and attractive perspectives},
	volume = {9},
	issn = {1878-5085},
	shorttitle = {Biobanks in {Horizon} 2020},
	url = {https://doi.org/10.1007/s13167-018-0153-7},
	doi = {10.1007/s13167-018-0153-7},
	language = {en},
	number = {4},
	urldate = {2022-01-07},
	journal = {EPMA Journal},
	author = {Kinkorová, Judita and Topolčan, Ondřej},
	month = dec,
	year = {2018},
	pages = {345--353},
}

@article{kinkorova_biobanks_2016,
	title = {Biobanks in the era of personalized medicine: objectives, challenges, and innovation},
	volume = {7},
	issn = {1878-5085},
	shorttitle = {Biobanks in the era of personalized medicine},
	url = {https://doi.org/10.1186/s13167-016-0053-7},
	doi = {10.1186/s13167-016-0053-7},
	language = {en},
	number = {1},
	urldate = {2022-01-07},
	journal = {EPMA Journal},
	author = {Kinkorová, Judita},
	month = feb,
	year = {2016},
	pages = {4},
}

@article{beesley_emerging_2020,
	title = {The emerging landscape of health research based on biobanks linked to electronic health records: {Existing} resources, statistical challenges, and potential opportunities},
	volume = {39},
	issn = {1097-0258},
	shorttitle = {The emerging landscape of health research based on biobanks linked to electronic health records},
	url = {https://onlinelibrary.wiley.com/doi/abs/10.1002/sim.8445},
	doi = {10.1002/sim.8445},
	language = {en},
	number = {6},
	urldate = {2021-11-29},
	journal = {Statistics in Medicine},
	author = {Beesley, Lauren J. and Salvatore, Maxwell and Fritsche, Lars G. and Pandit, Anita and Rao, Arvind and Brummett, Chad and Willer, Cristen J. and Lisabeth, Lynda D. and Mukherjee, Bhramar},
	year = {2020},
	note = {\_eprint: https://onlinelibrary.wiley.com/doi/pdf/10.1002/sim.8445},
	pages = {773--800},
}

@article{coppola_biobanking_2019,
	title = {Biobanking in health care: evolution and future directions},
	volume = {17},
	issn = {1479-5876},
	shorttitle = {Biobanking in health care},
	url = {https://doi.org/10.1186/s12967-019-1922-3},
	doi = {10.1186/s12967-019-1922-3},
	language = {en},
	number = {1},
	urldate = {2021-11-29},
	journal = {Journal of Translational Medicine},
	author = {Coppola, Luigi and Cianflone, Alessandra and Grimaldi, Anna Maria and Incoronato, Mariarosaria and Bevilacqua, Paolo and Messina, Francesco and Baselice, Simona and Soricelli, Andrea and Mirabelli, Peppino and Salvatore, Marco},
	month = may,
	year = {2019},
	pages = {172},
}

@article{garcia-merino_spanish_2009,
	title = {The {Spanish} {HIV} {BioBank}: a model of cooperative {HIV} research},
	volume = {6},
	issn = {1742-4690},
	shorttitle = {The {Spanish} {HIV} {BioBank}},
	url = {https://doi.org/10.1186/1742-4690-6-27},
	doi = {10.1186/1742-4690-6-27},
	language = {en},
	number = {1},
	urldate = {2021-11-29},
	journal = {Retrovirology},
	author = {García-Merino, Isabel and de las Cuevas, Natividad and Jiménez, José Luis and Gallego, Jorge and Gómez, Coral and Prieto, Cristina and Serramía, Ma Jesús and Lorente, Raquel and Muñoz-Fernández, Ma Ángeles and {Spanish HIV BioBank}},
	month = mar,
	year = {2009},
	pages = {27},
}

@article{ravid_standard_2008,
	title = {Standard {Operating} {Procedures}, ethical and legal regulations in {BTB} ({Brain}/{Tissue}/{Bio}) banking: what is still missing?},
	volume = {9},
	issn = {1573-6814},
	shorttitle = {Standard {Operating} {Procedures}, ethical and legal regulations in {BTB} ({Brain}/{Tissue}/{Bio}) banking},
	url = {https://doi.org/10.1007/s10561-007-9055-y},
	doi = {10.1007/s10561-007-9055-y},
	language = {en},
	number = {2},
	urldate = {2021-11-29},
	journal = {Cell and Tissue Banking},
	author = {Ravid, Rivka},
	month = jun,
	year = {2008},
	pages = {121--137},
}

@article{bradbury_diet_2020,
	title = {Diet and colorectal cancer in {UK} {Biobank}: a prospective study},
	volume = {49},
	issn = {0300-5771},
	shorttitle = {Diet and colorectal cancer in {UK} {Biobank}},
	url = {https://doi.org/10.1093/ije/dyz064},
	doi = {10.1093/ije/dyz064},
	number = {1},
	urldate = {2021-11-29},
	journal = {International Journal of Epidemiology},
	author = {Bradbury, Kathryn E and Murphy, Neil and Key, Timothy J},
	month = feb,
	year = {2020},
	pages = {246--258},
}

@article{kaye_single_2011,
	title = {From single biobanks to international networks: developing e-governance},
	volume = {130},
	issn = {1432-1203},
	shorttitle = {From single biobanks to international networks},
	url = {https://doi.org/10.1007/s00439-011-1063-0},
	doi = {10.1007/s00439-011-1063-0},
	language = {en},
	number = {3},
	urldate = {2021-11-29},
	journal = {Human Genetics},
	author = {Kaye, Jane},
	month = jul,
	year = {2011},
	pages = {377},
}

@article{clayton_informed_2005,
	title = {Informed {Consent} and {Biobanks}},
	volume = {33},
	issn = {1073-1105, 1748-720X},
	url = {https://www.cambridge.org/core/journals/journal-of-law-medicine-and-ethics/article/abs/informed-consent-and-biobanks/EE5D9287DE9ACCC6E1D9616130A942F4},
	doi = {10.1111/j.1748-720X.2005.tb00206.x},
	language = {en},
	number = {1},
	urldate = {2021-11-29},
	journal = {Journal of Law, Medicine \& Ethics},
	author = {Clayton, Ellen Wright},
	year = {2005},
	note = {Publisher: Cambridge University Press},
	pages = {15--21},
}

@article{cambon-thomsen_social_2004,
	title = {The social and ethical issues of post-genomic human biobanks},
	volume = {5},
	copyright = {2004 Nature Publishing Group},
	issn = {1471-0064},
	url = {https://www.nature.com/articles/nrg1473},
	doi = {10.1038/nrg1473},
	language = {en},
	number = {11},
	urldate = {2021-11-29},
	journal = {Nature Reviews Genetics},
	author = {Cambon-Thomsen, Anne},
	month = nov,
	year = {2004},
	note = {Bandiera\_abtest: a
Cg\_type: Nature Research Journals
Number: 11
Primary\_atype: Comments \& Opinion
Publisher: Nature Publishing Group},
	pages = {866--873},
}

@article{abbott_sweden_1999,
	title = {Sweden sets ethical standards for use of genetic ‘biobanks’},
	volume = {400},
	copyright = {1999 Macmillan Magazines Ltd.},
	issn = {1476-4687},
	url = {https://www.nature.com/articles/21720},
	doi = {10.1038/21720},
	language = {en},
	number = {6739},
	urldate = {2021-11-29},
	journal = {Nature},
	author = {Abbott, Alison},
	month = jul,
	year = {1999},
	note = {Bandiera\_abtest: a
Cg\_type: Nature Research Journals
Number: 6739
Primary\_atype: News
Publisher: Nature Publishing Group},
	pages = {3--3},
}

@article{greely_uneasy_2007,
	title = {The {Uneasy} {Ethical} and {Legal} {Underpinnings} of {Large}-{Scale} {Genomic} {Biobanks}},
	volume = {8},
	url = {https://doi.org/10.1146/annurev.genom.7.080505.115721},
	doi = {10.1146/annurev.genom.7.080505.115721},
	number = {1},
	urldate = {2021-11-29},
	journal = {Annual Review of Genomics and Human Genetics},
	author = {Greely, Henry T.},
	year = {2007},
	pmid = {17550341},
	note = {\_eprint: https://doi.org/10.1146/annurev.genom.7.080505.115721},
	pages = {343--364},
}

@article{rothstein_expanding_2005,
	title = {Expanding the {Ethical} {Analysis} of {Biobanks}},
	volume = {33},
	issn = {1073-1105, 1748-720X},
	url = {https://www.cambridge.org/core/journals/journal-of-law-medicine-and-ethics/article/abs/expanding-the-ethical-analysis-of-biobanks/61EB57A4AE1B08A4662D392ED67D8E55},
	doi = {10.1111/j.1748-720X.2005.tb00213.x},
	language = {en},
	number = {1},
	urldate = {2021-11-29},
	journal = {Journal of Law, Medicine \& Ethics},
	author = {Rothstein, Mark A.},
	year = {2005},
	note = {Publisher: Cambridge University Press},
	pages = {89--101},
}

@article{baker_building_2012,
	title = {Building better biobanks},
	volume = {486},
	copyright = {2012 Nature Publishing Group, a division of Macmillan Publishers Limited. All Rights Reserved.},
	issn = {1476-4687},
	url = {https://www.nature.com/articles/486141a},
	doi = {10.1038/486141a},
	language = {en},
	number = {7401},
	urldate = {2021-11-29},
	journal = {Nature},
	author = {Baker, Monya},
	month = jun,
	year = {2012},
	note = {Bandiera\_abtest: a
Cg\_type: Nature Research Journals
Number: 7401
Primary\_atype: Special Features
Publisher: Nature Publishing Group
Subject\_term: Diseases;Molecular biology
Subject\_term\_id: diseases;molecular-biology},
	pages = {141--146},
}

@article{hansson_ethics_2009,
	title = {Ethics and biobanks},
	volume = {100},
	copyright = {2009 The Author(s)},
	issn = {1532-1827},
	url = {https://www.nature.com/articles/6604795},
	doi = {10.1038/sj.bjc.6604795},
	language = {en},
	number = {1},
	urldate = {2021-11-29},
	journal = {British Journal of Cancer},
	author = {Hansson, M. G.},
	month = jan,
	year = {2009},
	note = {Bandiera\_abtest: a
Cg\_type: Nature Research Journals
Number: 1
Primary\_atype: Reviews
Publisher: Nature Publishing Group},
	pages = {8--12},
}

@article{hewitt_defining_2013,
	title = {Defining {Biobank}},
	volume = {11},
	issn = {1947-5535, 1947-5543},
	url = {http://www.liebertpub.com/doi/10.1089/bio.2013.0042},
	doi = {10.1089/bio.2013.0042},
	language = {en},
	number = {5},
	urldate = {2021-11-29},
	journal = {Biopreservation and Biobanking},
	author = {Hewitt, Robert and Watson, Peter},
	month = oct,
	year = {2013},
	pages = {309--315},
}

@article{bravo_developing_2015,
	title = {Developing a guideline to standardize the citation of bioresources in journal articles ({CoBRA})},
	volume = {13},
	issn = {1741-7015},
	url = {https://doi.org/10.1186/s12916-015-0266-y},
	doi = {10.1186/s12916-015-0266-y},
	number = {1},
	urldate = {2021-10-20},
	journal = {BMC Medicine},
	author = {Bravo, Elena and Calzolari, Alessia and De Castro, Paola and Mabile, Laurence and Napolitani, Federica and Rossi, Anna Maria and Cambon-Thomsen, Anne},
	month = feb,
	year = {2015},
	pages = {33},
}

@article{cambon-thomsen_role_2011,
	title = {The role of a bioresource research impact factor as an incentive to share human bioresources},
	volume = {43},
	copyright = {2011 Nature Publishing Group, a division of Macmillan Publishers Limited. All Rights Reserved.},
	issn = {1546-1718},
	url = {https://www.nature.com/articles/ng.831},
	doi = {10.1038/ng.831},
	language = {en},
	number = {6},
	urldate = {2021-10-20},
	journal = {Nature Genetics},
	author = {Cambon-Thomsen, Anne and Thorisson, Gudmundur A and Andrieu, Sandrine and Bertier, Gabrielle and Boeckhout, Martin and Cambon-Thomsen, Anne and Carpenter, Jane and Dagher, Georges and Dalgleish, Raymond and Deschênes, Mylène and di Donato, Jeanne Hélène and Filocamo, Mirella and Goldberg, Marcel and Hewitt, Robert and Hofman, Paul and Kauffmann, Francine and Leitsalu, Liis and Lomba, Irene and Mabile, Laurence and Melegh, Bela and Metspalu, Andres and Miranda, Lisa and Napolitani, Federica and Oestergaard, Mikkel Z and Parodi, Barbara and Pasterk, Markus and Reiche, Acacia and Rial-Sebbag, Emmanuelle and Rivalle, Guillaume and Rochaix, Philippe and Susbielle, Guillaume and Tarasova, Linda and Thomsen, Mogens and Thorisson, Gudmundur A and Zawati, Ma'n H and Zins, Marie and Mabile, Laurence and {the BRIF workshop group} and {Named collaborators}},
	month = jun,
	year = {2011},
	note = {Bandiera\_abtest: a
Cg\_type: Nature Research Journals
Number: 6
Primary\_atype: Correspondence
Publisher: Nature Publishing Group
Subject\_term: Computational biology and bioinformatics;Research data
Subject\_term\_id: computational-biology-and-bioinformatics;research-data},
	pages = {503--504},
}

@article{cambon-thomsen_assessing_2003,
	title = {Assessing the impact of biobanks},
	volume = {34},
	copyright = {2003 Nature Publishing Group},
	issn = {1546-1718},
	url = {https://www.nature.com/articles/ng0503-25b},
	doi = {10.1038/ng0503-25b},
	language = {en},
	number = {1},
	urldate = {2021-09-30},
	journal = {Nature Genetics},
	author = {Cambon-Thomsen, Anne},
	month = may,
	year = {2003},
	note = {Bandiera\_abtest: a
Cg\_type: Nature Research Journals
Number: 1
Primary\_atype: Correspondence
Publisher: Nature Publishing Group},
	pages = {25--26},
}
\end{refsection}
\end{document}


\maketitle

\section{Biobank sample}\label{SI:dataset}

In order to get a biobank sample, we looked for specific keywords in the titles of articles from the Microsoft Academic Graph (MAG, snapshot of August 2021). Based on a set of articles containing specific Medical Subject Headers (MeSH, See bellow), several categories of articles were then selected based on the keywords presented bellow. All articles should furthermore contain a colon (:) at some place in the title as they usually appear in the titles of articles introducing a biobank. Each set of articles obtained from the first filter was processed independently as explained below.

\subsection{MeSH Qualifier List}
We use the following keywords to filter only biomedical articles relevant for biobank studies. MeSH has two kinds of identifiers: Descriptors (main headings) and Qualifiers (subheadings). Here we based our filtering only at a subheading level, based on the following Qualifiers.

`Cohort Studies',
`Prospective Studies',
`Risk Factors',
`Longitudinal Studies',
`Research Design', 
`Surveys and Questionnaires', 
`Health Status',
`Follow-Up Studies',
`Chronic Disease', 
`Registries',
`Health Surveys',
`Prevalence', 
`Life Style', 
`Biomarkers'
`Cross-Sectional Studies', 
`Biological Specimen Banks', 
`Data Collection',
`Quality of Life',
`Health Behavior',
`Population Surveillance', 
`Patient Selection' and
`Multicenter Studies as Topic'.

\subsection{Cohort Baseline}
Articles with the words cohort and baseline were processed first by removing all titles with the words baseline study, baseline findings, and baseline analysis; as they usually signal a paper with results using the baseline data of a cohort instead of the introduction of a biobank. Articles selected also have to contain any of the following words: design, rationale, methods, characteristics, and objectives, as they could imply an introduction to a cohort.

\subsection{Health Study}
As with the titles including the keywords ``cohort'' and ``baseline'', titles with the keyword ``health study'' have to include the keywords design, rationale, methods, characteristics, and objectives. Furthermore, we selected only the titles starting with the word "The" to include only articles introducing a health study.

\subsection{Biobank}
We recovered two sets of articles based on whether they had the word "biobank" or "mega biobank" in their title. The former was processed in two parts separately, first by taking only the titles that started with the word "The", and the secondly by taking the titles that had only one word followed by the word "biobank" (e.g. Quatar Biobank). The two obtained sets of articles where then merges into a single set and combined with the list of titles contained the word "mega biobank".

\subsection{Cohort profile}
We considered all titles with the words "Cohort profile:" in their title as they usually indicate the presentation of a new or existing cohort profile to the scientific community. This includes the cohort profiles of the International Journal of Epidemiology and the BMC Open journals.

\subsection{Epidemiological/Epidemiology study}
We considered all titles with the words "Epidemiology" or "Epidemiological" and "study" in their title. Furthermore, only titles introducing a design or an approach are considered using the keywords: approach, design, method, rationale, characteristics, objectives, and approach. Finally, we recover titles starting with ``The [word] Study:'', or finishing with ``: The  [word] Study'' where  [word] is any alphanumerical word.

\subsection{Prospective study}
In order to obtain the articles introducing the design and rationale of prospective projects, we searched for titles with the word `prospective', `study', `rationale', and `design'.

\subsection{Longitudinal cohorts}
In order to obtain the articles introducing the design and rationale of longitudinal cohorts, we searched for titles with the word `longitudinal', `study', `rationale', and `design'.

\subsection{Cohort study}

In order to get the articles presenting a cohort study we looked for titles with the word `study', `design', `rationale', and `design'.

\subsection{Follow-up study}
Some population can span several biobanks depending on the follow-up data they may get. In order to include some of these biobanks in our cohort, we searched for the keywords `follow-up’, `rationale’, `design’ in the titles of articles.

\subsection{Manual Cleaning}

A final manual curation of the dataset is done to remove articles that fall in at least one of the following:

\begin{enumerate}
    \item Is not the first article available that introduces the biobank to the research community.
    \item Is not human based.
    \item Is not a research article.
    \item Is not related to a biobank.
\end{enumerate}

\section{Community Detection}\label{SI:communities}

The biobank co-citation network, which nodes are biobank papers that are connected by an edge if there is at least one article citing both biobanks, allows for the relational study of biobanks. This network has 540 nodes and 3,262 edges. Using the minimizing block model algorithm from \texttt{graph-tool}, a library in Python, we obtain six communities of biobanks.

In order to label the different communities by topic of study, we consider all the citations to the 540 biobanks in the network. Once we obtain the citations, we start by looking at the common words in the titles (Figure~\ref{fig:word-cloud}). Looking at the common words in each community we can broadly categorize the five communities by topic: population-based biobanks (purple in Figure~\ref{fig:word-cloud}), aging biobanks (brown), British biobanks (ocean green), developing nations biobanks (red), and birth/childhood biobanks (light green).

\begin{figure}
    \centering
    \includegraphics[width=1\linewidth]{word-cloud.pdf}
    \caption{\textbf{The co-citation network and its communities.} Each community is marked by a different color. Word clouds are added containing the keywords of the articles using the biobanks in each community.}
    \label{fig:word-cloud}
\end{figure}

We apply a generative probabilistic algorithm to the title corpus belonging to the citing papers called Latent Dirichlet Allocation (LDA)~\cite{blei_latent_2003}, obtaining the main research topics of each community.

{\footnotesize
\begin{landscape}

\begin{tabular}{llc}

\toprule
         Community &                                                                                                       Keywords &  Number of papers \\
\midrule
             aging &                          old, adult, longitudinal, population, base, aged, evidence, year, symptom, depressive &                     6047 \\
             aging &          cognitive, physical, activity, man, impairment, incident, disability, disorder, time, cross\_sectional &                      139 \\
             aging &                 association, life, function, result, difference, model, depression, sex, relate, socioeconomic &                      112 \\
             aging &                chinese, analysis, disease, status, cardiovascular, diabetes, dietary, index, urban, individual &                       93 \\
             aging &                                    woman, australian, finding, young, survey, study, body, brain, weight, cost &                       72 \\
   birth/childhood &                      birth, maternal, pregnancy, age, year, trajectory, offspre, outcome, finding, gestational &                     3530 \\
   birth/childhood &           childhood, early, analysis, adolescence, problem, parental, relate, impact, developmental, potential &                     1210 \\
   birth/childhood &                           young, genetic, prospective, adult, approach, trait, obesity, dietary, model, intake &                      663 \\
   birth/childhood &               child, longitudinal, parent, study, avon, function, alspac, psychopathology, adversity, european &                      547 \\
   birth/childhood &                     health, exposure, prenatal, infant, development, mental, postnatal, research, assess, link &                      285 \\
           british &                 health, social, adult, child, socioeconomic, old, physical, development, finding, relationship &                     1995 \\
           british &                                         birth, age, british, risk, early, disease, cohort, factor, brain, role &                      936 \\
           british &                         life, cognitive, associate, course, later, research, education, ability, late, predict &                      195 \\
           british &                    childhood, evidence, change, work, mid, problem, mobility, whitehall\_ii, ethnic, experience &                      106 \\
           british &                      study, longitudinal, population, base, outcome, impact, sample, swedish, prevalence, care &                       55 \\
developing country &       birth, child, year, analysis, pelota, association, longitudinal, adolescence, physical\_activity, pattern &                      638 \\
developing country &                               base, population, maternal, cohort, body, pregnancy, weight, mental, city, score &                      240 \\
developing country &                  early, childhood, young, life, growth, disease, effect, cardiovascular, adulthood, guatemalan &                      206 \\
developing country &                   health, adolescent, evidence, study, outcome, nutrition, cognitive, profile, preterm, result &                      138 \\
developing country &                         adult, south, african, status, relation, urban, trajectory, nutritional, black, marker &                      106 \\
      health based &                  risk, patient, population, mortality, korean, association, change, year, increase, prevalence &                     2857 \\
      health based &                         treatment, use, therapy, antiretroviral, control, trend, disorder, follow, large, loss &                      875 \\
      health based &                          disease, chronic, kidney, cric, progression, index, finding, death, prediction, rural &                      788 \\
      health based &                         diabetes, type, prevention, program, primary, research, method, screen, art, fremantle &                      291 \\
      health based &                             live, people, level, survivor, childhood, swiss, social, database, child, hospital &                      182 \\
  population based &                                 adult, woman, old, result, year, consumption, intake, function, index, outcome &                     9149 \\
  population based &                          cancer, mortality, prospective, breast, specific, dietary, sex, cause, biomarker, man &                     3471 \\
  population based &                            association, base, wide, genome, sleep, identify, score, individual, obesity, model &                     3438 \\
  population based & disease, cardiovascular, gene, mendelian\_randomization, level, relationship, stroke, circulate, marker, causal &                     1999 \\
  population based &                     population, diabetes, type, datum, predict, incident, general, research, pattern, behavior &                     1748 \\
\bottomrule
\end{tabular}

\end{landscape}
}

\section{Diseases studied}\label{SI:diseases}

\subsection{Most studied diseases}

\section{Biobank $h$-index}\label{SI:biobank-h-index}

\section{Biobank Affiliation}\label{SI:affiliations}
Given the list of authors of a biobank, we define the country of the biobank as the country of affiliation of its PI. If the PI has more than one affiliation, we take the mode of the country list from all the authors of the biobank paper. From the original list of 1,209 biobanks, we obtained the affiliation of their lead author for 710 of them. Similarly, we take the country of the affiliation of the lead author to obtain the origin of the citing papers. From 58,277 citations to biobanks, we identify the country of 40,296 articles.

\section{Null model}\label{SI:null-model}

\section{Authors}\label{SI:authors}

For each biobank paper, we recover all its references and citations and call create two new sets of papers. For each biobank, citation and reference paper we then look for all the authors involved in each paper. This gives us a list of authors for each article, together with the affiliation(s) of the author (at the time of the paper), and its sequence number in the co-authorship. The leading scientist of a biobank is defined as the last author in the biobank paper and we take the country of the biobank to be equal to the author affiliation's country at the time of the article. For those authors without affiliation at the time of the paper, we take their last-known affiliation, and define the country of the biobank as the country of this affiliation.

We recover the authors of 1,163 out of 1,209 papers in the original biobank list of articles. From those, 944 have the affiliation of the leading scientist and 939 the country of the affiliation. The same is done regarding the origins of citing papers: we took the affiliation's country of each leading scientist. From the original 55,850 citing papers, we recovered the affiliation country of 47,496.

\section{Biobank countries}\label{SI:countries}
We identified the country of origin for 914 biobanks, extracted from the main affiliation of their authors (SI).  In total, biobanks come from 50 different countries, however, the distribution is far from balanced: the USA hosts 25\% of all biobanks, followed by the UK (16\%) and Australia (10\%). Altogether the top five countries host  62\% of all biobanks (SI).

We identified 55,850 unique papers citing a biobank, and identified the affiliation of 40,706  of their lead scientist. We find that biobank data is used in 108 countries, more than twice the number of countries hosting a biobank. Usage is also heavily concentrated on a small number of countries, i.e., USA (28.6\%), UK (15.6\%), and Australia (9\%), reflecting the distribution of global medical research.

Global north and south countries differ in terms of social-economic status, and are not necessarily within the geographical names~\cite{hollington_concepts_2015}. The categories used in the country are taken from the Wikimedia foundation classification, which can be found at https://github.com/wikimedia-research/canonical-data/blob/master/countries.csv.

\section{Global south biobanks not led by global south scientists}
Biomedical research is hardly a one-person job. Data-based research spurs collaboration, and cutting-edge research in life sciences tends to come from well-funded and large teams~\cite{simonton_scientific_2013}. Indeed, the average biobank team in our sample is composed of 12.3 individuals. As team sizes increase, who gets the credit becomes an important question~\cite{shen_collective_2014}, especially for international collaborations between developed and developing countries, where credit imbalance may arise~\cite{serwadda_open_2018}. Here we look at the regional distribution of authors in both the origin and usage of biobanks.

Global south scientists are found in 225 out of the 1,217 biobanks from which we could extract the affiliation of their members, including high-impact biobanks like The China Health and Nutrition Survey with 486 citations, and the four \textit{Pelotas} studies from Brazil (1982, 1993, 2004, and 2015), summing together 610 citations. Most biobanks are composed by teams exclusively from the global north\footnote{High income countries, usually from the geographic north, except for Australia and New Zealand.}, and the general composition of biobanks with global south scientists remains an open question.

The ``90/10 gap'', coined by the Commission on Health Research for Development refers to the fact that 90\% of the biomedical research focuses on diseases affecting only 10\% of the world population~\cite{development_health_1990}. We find that biobank use and origin follows a similar pattern: 90\% of the biobanks are based in the global north, which also produces 91\% of the research relying on biobanks. One exception is the community of biobanks from developing countries, with more than half employing a member from the global south, and 46\% of their impact coming from there.

From the 31 biobanks from developing countries, only 7 are led by a scientist in the global south. A similar situation occurs within the teams producing research with their data: even though the majority contain at least one author from the global south, only 36\% are led by an author from the region. An effect called helicopter research, where scientist from the global north publish results with little involvement in key positions from local collaborators~\cite{serwadda_open_2018,haelewaters_ten_2021}. In general, only 30\% of the articles using data from developing and middle income countries have a first and last author in the global south, as opposed to 56\% from the global north.

\section{Hidden citations}\label{hidden-citations}

We qualify a citation as ``hidden'' if a citing paper uses the biobank name in its abstract but does not cite the biobank representative paper. We only considered articles that appeared after the introductory paper of the biobank and therefore could have the possibility to cite it.

\section{Logistic regression models}\label{SI:ML}

\section{Variability of biobanks}

Biobanks are repositories of biological specimens such as dried blood, brain or nails~\cite{chen_china_2011,ravid_standard_2008}, but may also contain digital data in the form of health information~\cite{garcia-merino_spanish_2009,fried_cardiovascular_1991}, diet intake~\cite{bradbury_diet_2020}, and DNA Sequencing~\cite{gaziano_million_2016,krokstad_cohort_2013}. They are used to better understand the relation between human traits and a broad range of conditions, like HIV~\cite{the_swiss_hiv_cohort_study_cohort_2010,tanser_cohort_2008}, ageing~\cite{bennett_rush_2005}, diabetes~\cite{moffet_cohort_2009}, cardiovascular disease~\cite{fried_cardiovascular_1991}, respiratory disease~\cite{marbury_indoor_1996}, psychiatric disease~\cite{otoole_australian_1996}, among others. They target populations of different ages including children~\cite{repka_cataract_2016}, teenagers~\cite{copeland_characteristics_2011}, adults~\cite{arnold_cohort_2019}, and the elderly~\cite{sachdev_sydney_2010}; as well as women~\cite{carlson_maine_1994} and men~\cite{norman_cohort_2009} cohorts. Access to data necessary for research is the central goal of biobanks, and they are unavoidable in epidemiological studies such as genome-wide analysis (GWAS) or single nucleotide polymorphism (SNPs) studies.

The inherent properties of biobanks vary greatly. Some biobanks target particular phenotypes and diseases, while others encompass a wide array of chronic diseases and are designed for general purpose~\cite{coppola_biobanking_2019}. Their specimen samples may come from a health system, such as a hospital or a health-care institution; or can be sampled from the population of a city or country, and the number of people sampled varies from a few individuals to hundreds of thousands~\cite{beesley_emerging_2020}. They are present all over the world and the data provided by them is used across many research institutions. This strong variability could explain, in part, their large impact variability: while some biobanks are used by a single research group, others help advance the research of thousands of scientists. How broad the impact of biobanks is, has been difficult to measure, as their use is not easy to track and quantitative studies of their impact are hard to find~\cite{odonoghue_how_2021}. 


The meta analysis of biobanks themselves has been of interest for ethical and methodological perspectives~\cite{hewitt_defining_2013}, and has spanned more than two decades of legal and technological transformation~\cite{hansson_ethics_2009,baker_building_2012,gaskell_biobanks_2011,rothstein_expanding_2005,greely_uneasy_2007,abbott_sweden_1999}. Ethical issues regarding participation consent and genetic data usage have been thoroughly explored~\cite{cambon-thomsen_social_2004}, as well as the best infrastructure methodologies to store, maintain and share biological samples and data~\cite{clayton_informed_2005,kaye_single_2011}. More recently, a biobank impact factor was proposed to track biobank data usage and impact in science~\cite{cambon-thomsen_role_2011}. This would allow for the quantification of the academic recognition gained by biobank teams, which in actuality is poorly understood, and is even expected to be lacking~\cite{cambon-thomsen_assessing_2003,bravo_developing_2015,mabile_quantifying_2013}.

\bibliography{zotero}
\bibliographystyle{plain}